\documentclass[11pt,a4paper]{emulateapj}
 \usepackage{url}
\usepackage{hyperref}
\usepackage{amsmath}
\usepackage[titletoc]{appendix}
\usepackage{xcolor}
\usepackage{graphicx}
\usepackage[export]{adjustbox}
\usepackage{psfig}
\usepackage[flushleft]{threeparttable}

\usepackage[symbol*]{footmisc}
\DefineFNsymbolsTM{myfnsymbols}{
 \textasteriskcentered *
  \textdagger    \dagger
  \textsection   \mathsection
  \textbardbl    \|%
  \textparagraph \mathparagraph 
}%
\setfnsymbol{myfnsymbols}
 \begin{document}

\title{gas sloshing regulates and records the evolution of the Fornax Cluster}

\author{Yuanyuan Su\altaffilmark{$\ddagger$1}}
\author{Paul E.\ J.\ Nulsen\altaffilmark{1}}
\author{Ralph P.\ Kraft\altaffilmark{1}}
\author{Elke Roediger\altaffilmark{2}}
\author{John A.\ ZuHone\altaffilmark{1}}
\author{Christine Jones\altaffilmark{1}}
\author{William R.\ Forman\altaffilmark{1}}
\author{Alex Sheardown\altaffilmark{2}}
\author{Jimmy A.\ Irwin\altaffilmark{3}}
\author{Scott W.\ Randall\altaffilmark{1}}
\affil{$^1$Harvard-Smithsonian Center for Astrophysics, 60 Garden Street, Cambridge, MA 02138, USA}
\affil{$^2$E.A. Milne Centre for Astrophysics, School of Mathematics and Physical Sciences, University of Hull, Hull, HU6 7RX, United Kingdom}
\affil{$^3$Department of Physics and Astronomy, University of Alabama, Box 870324, Tuscaloosa, AL 35487, USA}

\altaffiltext{$\ddagger$}{Email: yuanyuan.su@cfa.harvard.edu}


\begin{abstract}

We present results of a joint {\sl Chandra} and {\sl XMM-Newton} analysis of the Fornax Cluster, the nearest galaxy cluster in the southern sky. Signatures of merger-induced gas sloshing can be seen in the X-ray image. We identify four sloshing cold fronts in the intracluster medium, residing at radii of 3\,kpc (west), 10\,kpc (northeast), 30\,kpc (southwest) and 200\,kpc (east). Despite spanning over two orders of magnitude in radius, all four cold fronts fall onto the same spiral pattern that wraps around the BCG NGC~1399, likely all initiated by the infall of NGC~1404. The most evident front is to the northeast, 10\,kpc from the cluster center, which separates low-entropy high-metallicity gas and high-entropy low-metallicity gas. The metallicity map suggests that gas sloshing, rather than an AGN outburst, is the driving force behind the redistribution of the enriched gas in this cluster. The innermost cold front resides within the radius of the strong cool core. The sloshing time scale within the cooling radius, calculated from the Brunt-V$\ddot{\rm a}$s$\ddot{\rm a}$l$\ddot{\rm a}$ frequency, is an order of magnitude shorter than the cooling time. It is plausible that gas sloshing can contribute to the heating of the cool core, provided that gas of different entropies can be mixed effectively via Kelvin-Helmholtz instability. The estimated age of the outermost front suggests that this is not the first infall of NGC~1404.

\end{abstract}

\section{\bf Introduction}

A long-lasting puzzle in extragalactic astronomy is why 
galaxy clusters come in two varieties: cool-core clusters and non-cool-core clusters (Jones \& Forman 1984). The former feature a sharp X-ray emission peak, owing to the radiative cooling of a dense, cool, and enriched core.
{Gas in the cool core emits strongly in X-rays with a radiative cooling time much shorter than the age of the cluster. 
If there is no heating to compensate for radiative losses, a cooling flow with prodigious star formation is expected. High-resolution X-ray observations from {\sl Chandra} and {\sl XMM-Newton} have, however, revealed little low-temperature gas, posing the so called ``cooling problem" (see Peterson \& Fabian 2006 for a review).
Feedback from active galactic nuclei (AGN) may have provided the heat required to balance radiative cooling, despite the fact that the interplay between cooling and feedback remains a subject of debate.}
In contrast, the gaseous, thermal, and chemical distributions of non-cool-core clusters are relatively homogenous (Cavagnolo et al.\ 2009; Sanderson et al.\ 2009). 

A phenomenon called ``sloshing cold front" seems to be {\it exclusively} associated with cool-core clusters. 
Sloshing cold fronts produce characteristic spiral or bow-like features in the
X-ray surface brightness in cluster centers (Markevitch \& Vikhlinin 2007). Gas sloshing can be triggered by minor mergers or off-axis mergers that disturb the gas at the bottom of a cluster potential. The steep central entropy gradient in a cool-core cluster allows
sloshing to bring low-entropy central gas into contact with the higher
entropy intracluster medium (ICM), creating an abrupt step in the X-ray emissivity that
appears as a ``sloshing cold front"; this speculation has been confirmed by simulation (Ascasibar \& Markevitch 2006).

Whether sloshing can contribute to the suppression of cooling depends on the microscopic physics in the ICM --- how effectively heat can be transported.
Cold fronts are shear interfaces, i.e., the two layers of gas move with respect to each other parallel to the interface. In a purely hydrodynamical context, this configuration inevitably leads to the Kelvin-Helmholtz instability (KHI) (Chandrasekhar 1961; Lamb 1932), which promotes turbulent mixing between gas of different phases. 
KHI are often identified as ``roll" features just outside the interface in simulations and 
can, in principle, be observed directly as fine structures at the interface.
Deviations from pure hydro, in particular viscosity and magnetic fields, can suppress KHI and preserve the sharpness of the cold front (ZuHone et al.\ 2011; Roediger et al.\ 2013).
A growing number of deep {\sl Chandra} observations favor an inviscid ICM based on the presence of KHI eddies (Su et al.\ 2017a; Ichinohe et al.\ 2017).
Vikhlinin et al.\ (2001) demonstrate that it requires an ordered magnetic field of 10$\mu$G to suppress instability. This value is 10$\times$ larger than the strength of the intracluster magnetic field inferred from Faraday rotation and inverse Compton (Govoni \& Feretti 2004).
{It was suggested that a weak, tangled magnetic field can be stretched by shear in the sloshing gas to form a magnetic layer parallel to the cold front (Keshet et al.\ 2010).
The formation of such a field structure, called ``magnetic draping," can protect the front by inhibiting the growth of KHI (Dursi \& Pfrommer 2008; Lyutikov 2006).
This process has been reproduced in Magnetohydrodynamics (MHD) simulations (ZuHone et al.\ 2011).} 
If high and low-entropy gas on either side of the interface cannot be effectively mixed, then the heating of the core due to sloshing would be modest.

The atmospheres of galaxy clusters are observed to have elemental abundances that are
approximately 1/3 of the solar value ($Z_{\odot}$) over the bulk of their volume (0.3--1.0 virial radius), {which may have been enriched early on by supernovae and/or by the accumulation of the metals stripped from infalling galaxies} (e.g., De Grandi \& Molendi 2001; Dupke \& White 2000). Cluster centers ($r<0.3$\,virial radius) often show an enhanced metallicity of approximately $Z_{\odot}$. The excess metals are believed to be stellar ejecta from the brightest cluster galaxy (BCG). 
However, the extent of the $\simeq Z_\odot$ gas is often
greater than the optical extent of the BCG, causing a positive gradient in
the iron-mass-to-light ratio (e.g., David \& Nulsen 2008). This gradient implies
that some mechanism transports enriched central gas outward (Rebusco et al.\ 2005).
A number of observations of high-metallicity low-entropy gas surrounding buoyantly-rising AGN bubbles suggest that AGN outbursts are capable of displacing gas from the cluster center (e.g., Simionescu et al.\ 2009; Nulsen et al.\ 2002; Kirkpatrick \& McNamara 2015).
On the other hand, gas motion induced by mergers is also considered to be a mechanism to stir up the central gas. 
{In cool-core clusters, spatial correlations are found between the sloshing front and the elevated metallicities (Sanders et al.\ 2016a; Ghizzardi et al.\ 2014; 	
Blanton et al.\ 2011),} demonstrating that gas sloshing is capable of lifting a significant amount of metals from the cluster center. 
More generally, the role of mergers in redistributing enriched gas is manifested in an apparent association between a more disturbed cluster atmosphere and a more extended metallicity distribution (De Grandi \& Molendi 2001; Rossetti \& Molendi 2010; Su et al.\ 2016).

\begin{figure*}
   \centering
    \includegraphics[width=0.48\textwidth]{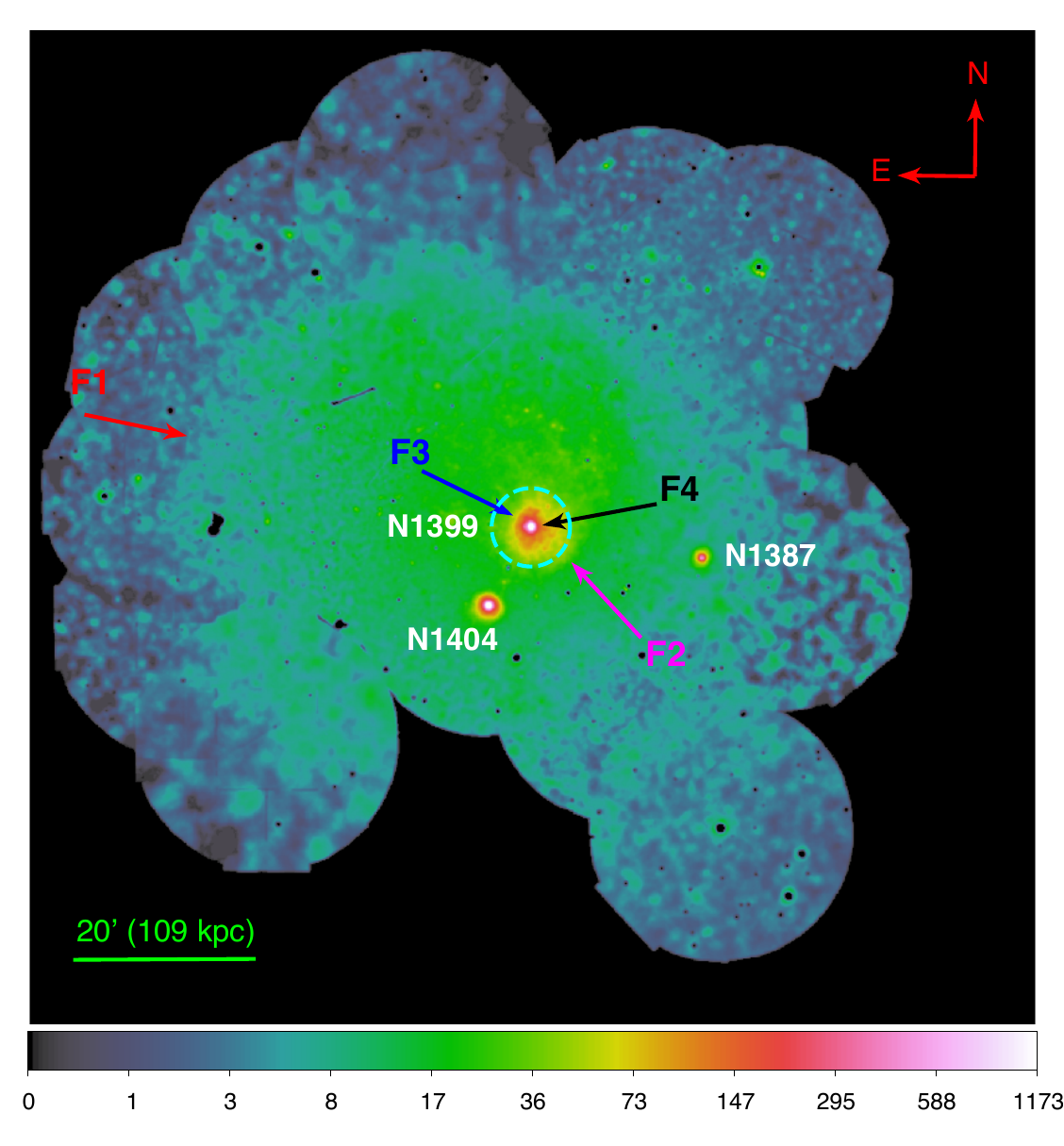}
        \includegraphics[width=0.48\textwidth]{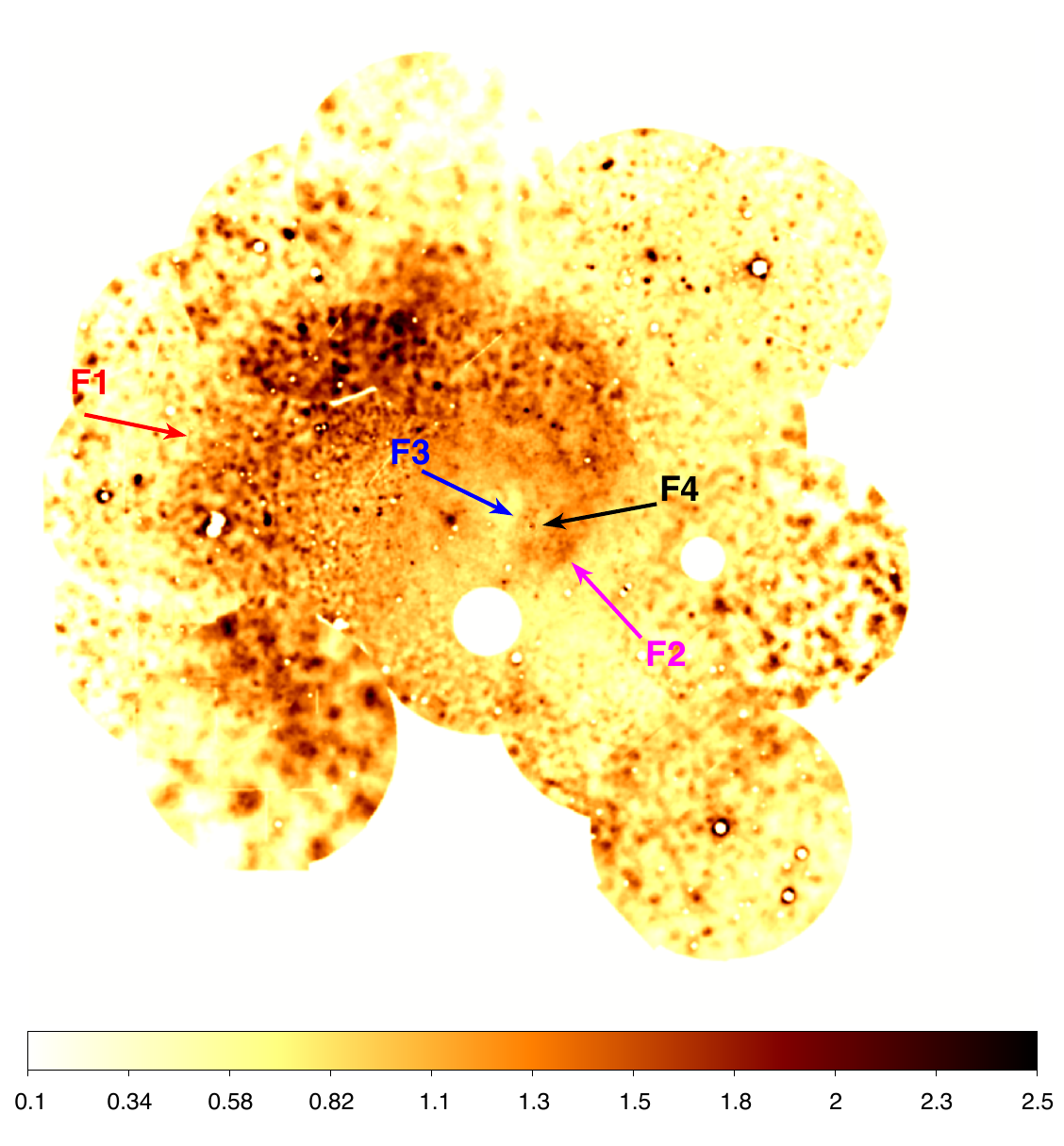}
\figcaption{\label{fig:fox} {\bf left}: {\sl XMM-Newton} mosaic image of the Fornax Cluster in the 0.5--2.0 keV energy band in units of cts$/$s$/$deg$^2$. This image is exposure- and vignetting- corrected with instrumental backgrounds subtracted. Point sources have been removed. Gas motions on large scales are suggested by the presence of multiple edges and a lack of spherical symmetry. The locations of the four sloshing cold fronts are indicated by the arrows. Cyan dashed circle: field-of-view of the {\sl Chandra} image in Figure~\ref{fig:img}. {\bf right}: The matching residual image of the left image with respect to the azimuthal average.}
\end{figure*}

In this paper, we present a case study of the nearest cluster in the southern sky, the Fornax Cluster ($D_L=19$\,Mpc, $1^{\prime} = 5.49$ kpc, $z=0.00475$, Su et al.\ 2017a,b), which is a low-mass cool-core cluster.
We show that the Fornax Cluster displays all the characteristics of a sloshing cluster, such as a spiral-shaped asymmetry (alternating pattern) in X-ray brightness, temperature, and metallicity. 
The BCG of this cluster, NGC~1399,
harbors a pair of symmetric radio lobes coincident with two X-ray cavities along a north-south axis.
Their properties are presented in a separate publication (Su et al.\ 2017c; hereafter Paper I).
Its cooling radius as a weak cool core (with a cooling time below 7.7\,Gyr) is 25\,kpc, while as a strong cool core (with a cooling time below 1\,Gyr) it is 4.5\,kpc.
{We found in NGC~1399 that cool gas uplifted by AGN bubbles can account for all of the gas that is expected to cool catastrophically, although the uplifted cool gas may eventually fall back.} 
Its second brightest galaxy, NGC~1404, is falling inward through the ICM from the southeast and features a sharp merging cold front and a stripped gaseous tail, which have been studied in detail in Machacek et al.\ (2005) and Su et al.\ (2017a,b). 
The large scale ICM has been observed extensively in the X-ray. 
{\sl ROSAT} observations indicate an average ICM temperature of $<1.5$\,keV (Rangarajan et al.\ 1995; Jones et al.\ 1997; Paolillo et al.\ 2002). A mosaic {\sl Chandra} observation with ten 50\,ksec pointings reveals an asymmetry in its morphology and temperature structures (Scharf et al.\ 2005).  
Using joint {\sl Suzaku} and {\sl XMM-Newton} mosaic observations, Murakami et al.\ (2011) 
find 
an average metallicity of near the solar value in the cluster center that declines to $\approx0.3Z_{\odot}$ at large radii, similar to that of many other galaxy clusters and groups. 
This study focuses on the effect of gas sloshing in the ICM. 
The observations and data reductions are described in \S2. Further analysis and the thermal and chemical properties are presented in \S3.
The implications of gas cooling and metal redistributions are discussed in \S4, and our main conclusions are summarized in \S5. Uncertainties reported are quoted at the 68\% confidence level throughout this work.

\section{\bf observations and data reductions}

\subsection{XMM-Newton}
We include all the existing {\sl XMM-Newton} observations within 1$^{\circ}$ of NGC~1399 as listed in Table~A1. Basic data reductions including screening and background modeling were performed using the Science Analysis System (SAS) version xmmsas-20160201.
All the ODF files were processed using {\tt emchain} and {\tt epchain} to ensure the latest calibrations. Soft flares were filtered from MOS data and pn data using the XMM-ESAS tools {\tt mos-filter} and {\tt pn-filter} respectively (Snowden \& Kuntz 2013). 
The effective exposure time of each detector is listed in Table~A1. The combined exposure time is $\approx$\,100\,ksec at the center and $\gtrsim$\,200\,ksec at the outskirts. 
We only include events files with FLAG\,$=0$ and PATTERN\,$<=12$ for MOS data and with FLAG\,$=0$ and PATTERN\,$<=4$ for pn data.    
Point sources detected by {\tt edetect\_chain} and confirmed by eye were excluded from further analysis. Out-of-time pn events were removed from both spectral and imaging analyses. Modeling of the astrophysical background (AXB) and the Non-X-ray background (NXB) is presented in the Appendix.  

Spectral fit of regions of interest was restricted to the 0.5--7.0 keV energy band. These spectra were fit to two sets of models simultaneously.
The first model set takes the form of {\tt phabs}$\times$({\tt pow}$_{\rm CXB}$+{\tt apec}$_{\rm MW}$+{\tt vapec}$_{\rm ICM}$)+{\tt apec}$_{\rm LB}$.  
Parameters of the AXB models were fixed to the values listed in Table~A3, which were determined with offset pointings.
The thermal {\tt vapec}$_{\rm ICM}$ model is for the cluster emission. 
The abundances of O, Ne, Mg, Si, S, Fe, and Ni were allowed to vary freely; all other elements were tied to Fe. 
The second model set is to characterize the NXB components (Table~A3) and their spectra were not folded through the Auxiliary Response Files (ARF). 
Parameters of the NXB models were fixed to the best-fits determined for each observation.

We created images in the
0.5--2.0 keV energy band. 
Individual detector images were created using the tasks
{\tt mos-spectra} and {\tt pn-spectra}. 
Point sources detected by the {\tt cheese} routine were removed. 
We used the XMM-ESAS tasks {\tt comb} and {\tt adapt\_900} to create
a background-subtracted, vignetting-corrected EPIC mosaic
image, binned by a factor of 2 and adaptively smoothed with a minimum of
50 counts per bin. The resulting image is shown in Figure~\ref{fig:fox}-left.

\subsection{Chandra}

\begin{figure*}
   \centering
    \includegraphics[width=0.45\textwidth]{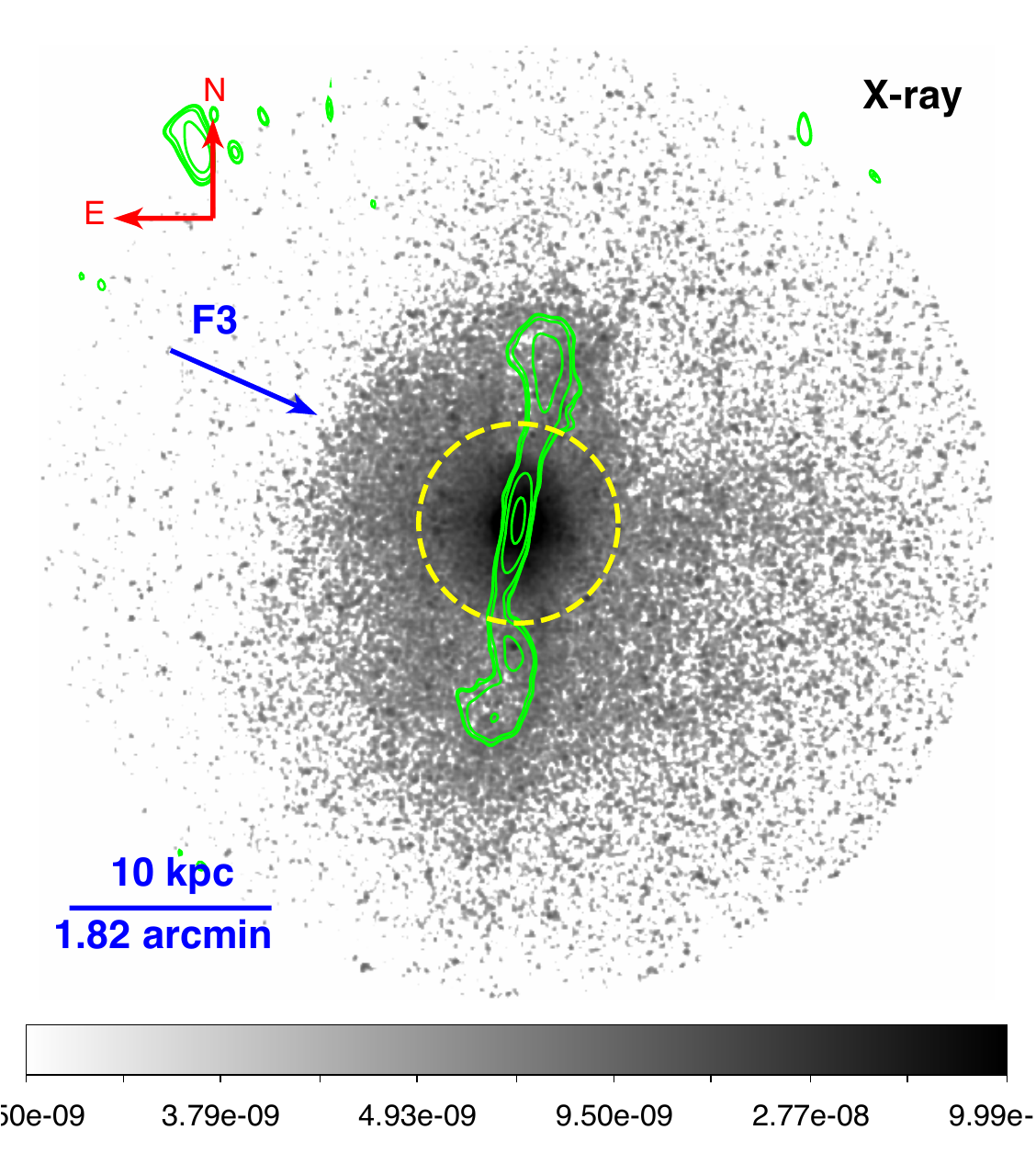}
                             \includegraphics[width=0.45\textwidth]{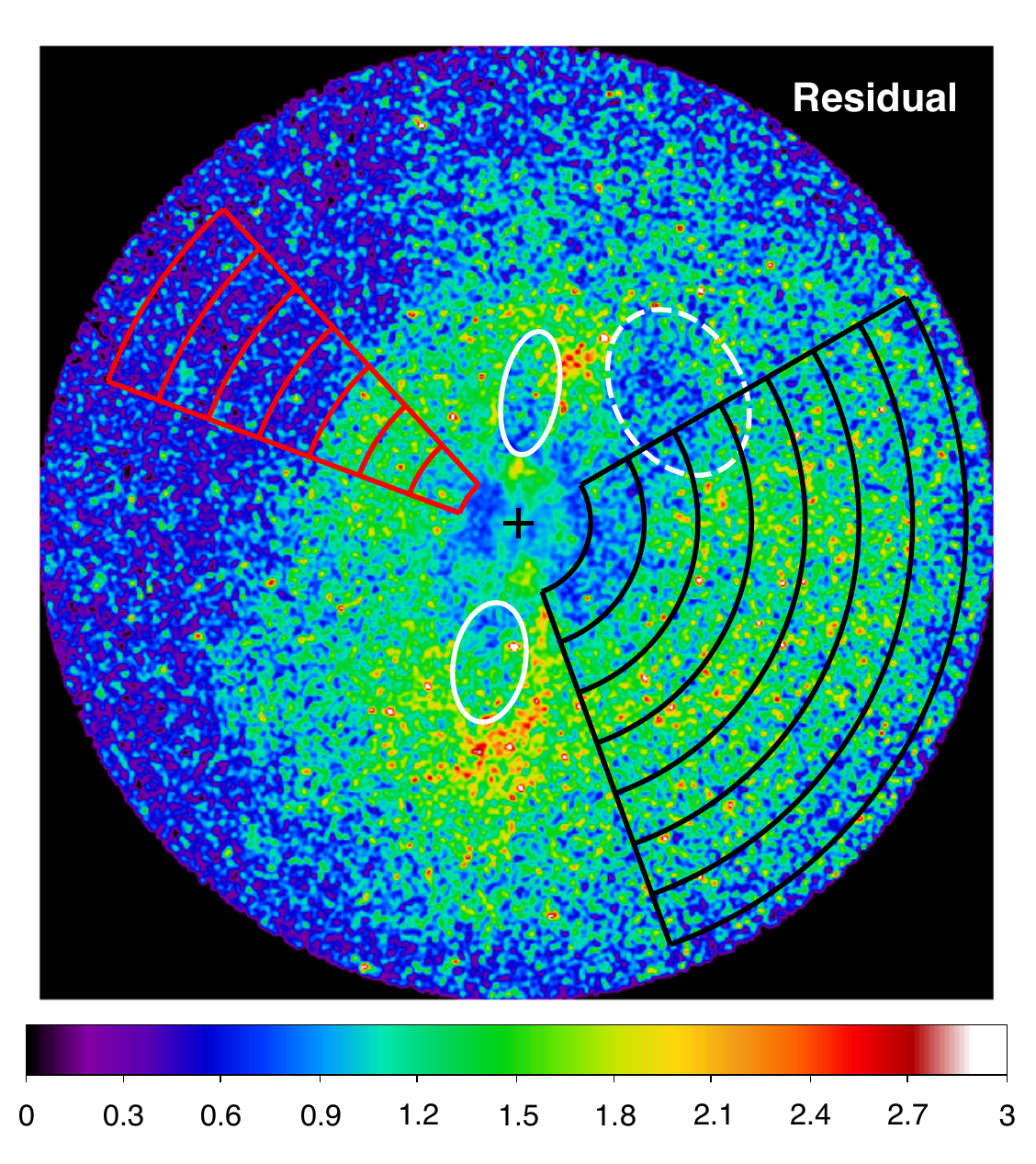} 
              \includegraphics[width=0.45\textwidth]{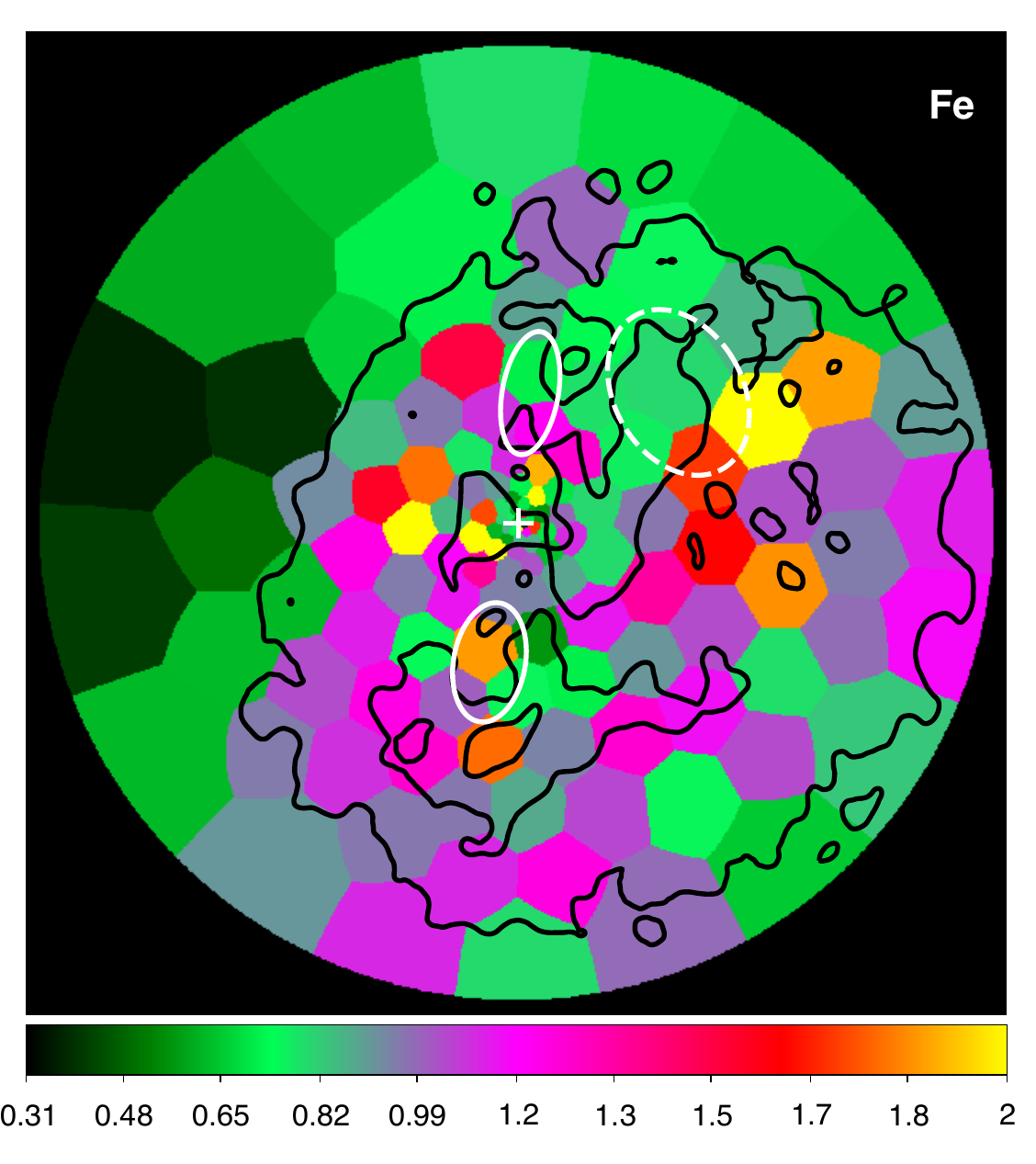}
                      \includegraphics[width=0.45\textwidth]{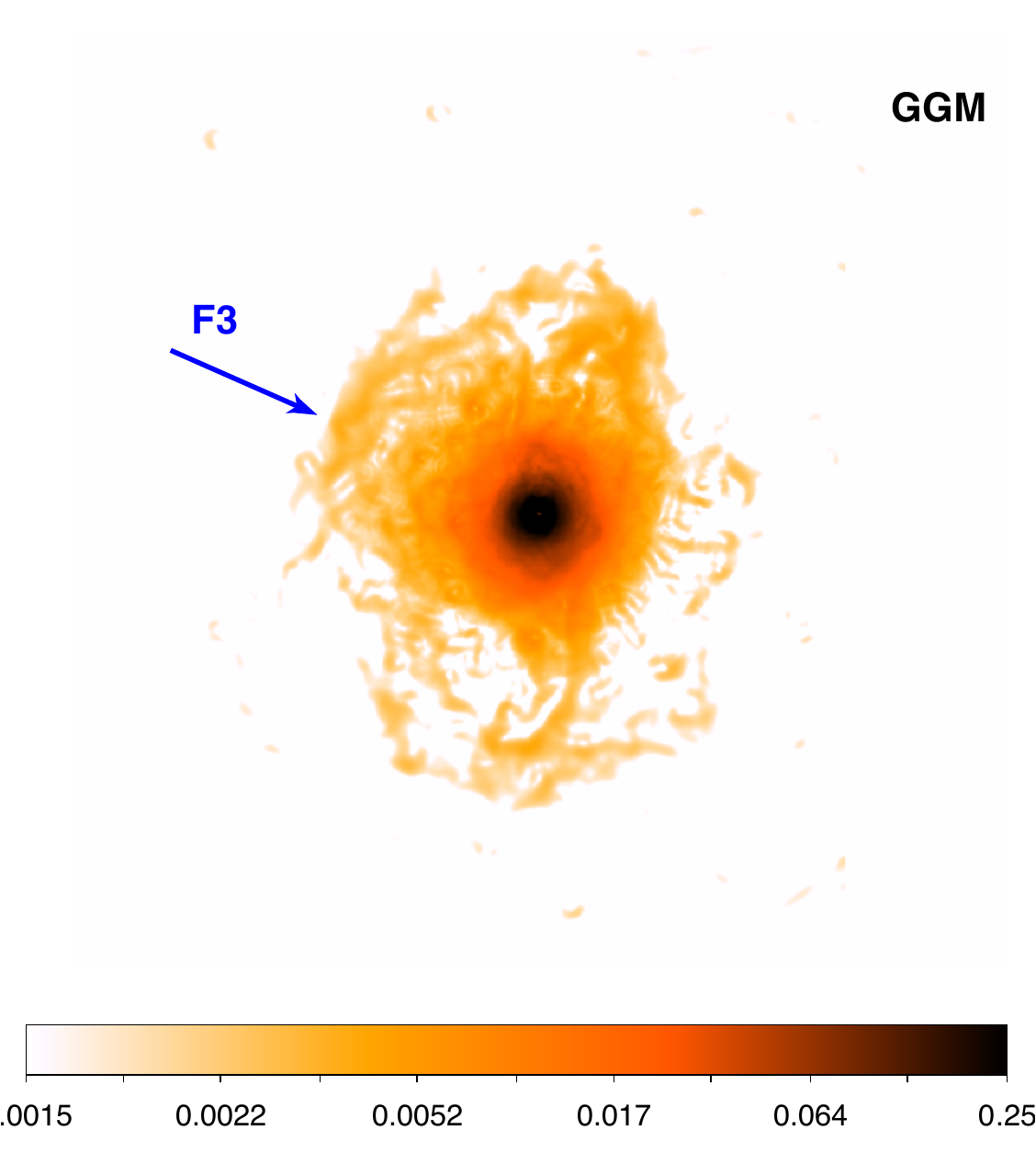} 
\figcaption{\label{fig:img} Gas properties within the weak cool core of Fornax ($r<25$\,kpc; $t_{\rm cool}<7.7$\,Gyr), corresponding to the cyan dashed circle in Figure~\ref{fig:fox}-left. {\bf top-left}: {\sl Chandra} X-ray image of NGC~1399 in the 0.5--2.0 keV energy band in units of photon\,cm$^{-2}$\,s$^{-1}$ per pixel ($0\farcs492\times0\farcs492$). The image was exposure-corrected and background-subtracted. The X-ray cavities extend north-south about the center to a radius of 10 kpc. Green contour: VLA 6 cm radio contour levels are set at [5, 5.4, 7.9, 22, 100]$\times$$\sigma_{\rm rms}$ where $\sigma_{\rm rms}$=0.1 mJy beam$^{-1}$. Yellow dashed circle: field-of-view of the {\sl Chandra} image of the inner structure in Figure~\ref{fig:img2}.
{\bf top-right}: The matching residual X-ray image, obtained by dividing the top-left image with a double $\beta$-model. The shapes of the bubbles are approximated by two white solid ellipses. The white dashed ellipse marks the position of a ghost cavity candidate.  
Black cross marks the cluster center (03h38m29s, -35d27m02s). 
{\bf bottom-left}: Two-dimensional Fe abundance distribution of the hot gas in NGC~1399 in units of solar abundance, derived with a two-temperature thermal model. White cross: cluster center. Black contours: the residual X-ray image (top-right). 
{\bf bottomr-right}: Gaussian Gradient Magnitude filtered image of the top-left image, obtained by combining images on scales of 2$\sigma$, 4$\sigma$, 8$\sigma$, 16$\sigma$, and 32$\sigma$.
}
\end{figure*}

We analyzed a combined of 250 ksec {\sl Chandra} observations centered on NGC~1399 (Obs-ID: 319, 4172, 9530, 14527, 14529, and 16639).
We refer interested readers to Paper 1 for details of the observations, data preparation, and the deprojected spectral analysis. 
All the data were reduced using {\sl CIAO}~4.9 and {\sl CALDB}~4.6.9 following standard procedures.  
We produced the blank-sky background subtracted, exposure corrected, and point source removed image which covers
the entire weak cool core ($r<25$\,kpc; $t_{\rm cool}<7.7$\,Gyr), as presented in Figure~\ref{fig:img}-top-left. 
In order to enhance detailed structures in the X-ray surface brightness, we divide this 0.5--2.0 keV image by its best-fit double-$\beta$ model. The resulting residual image is shown in Figure~\ref{fig:img}-top-right. 
Readout artifacts were subtracted in both imaging and spectral analyses.
Spectral fits were performed with {\sl XSPEC} 12.7 and C-statistics; the solar abundance standard of Asplund at al.\ (2006) was adopted. 
We use the thermal emission model ${\tt vapec}$ to model the hot gas component and a power law model with an index of 1.6, ${\tt pow}_{1.6}$, to describe the unresolved low mass X-ray binaries (LMXB) (Irwin et al.\ 2003). 
The deprojected temperature is below 1\,keV at the cluster center and rises to 1.5\,keV beyond 10\,kpc. 

\begin{figure*}
   \centering
       \includegraphics[width=0.45\textwidth]{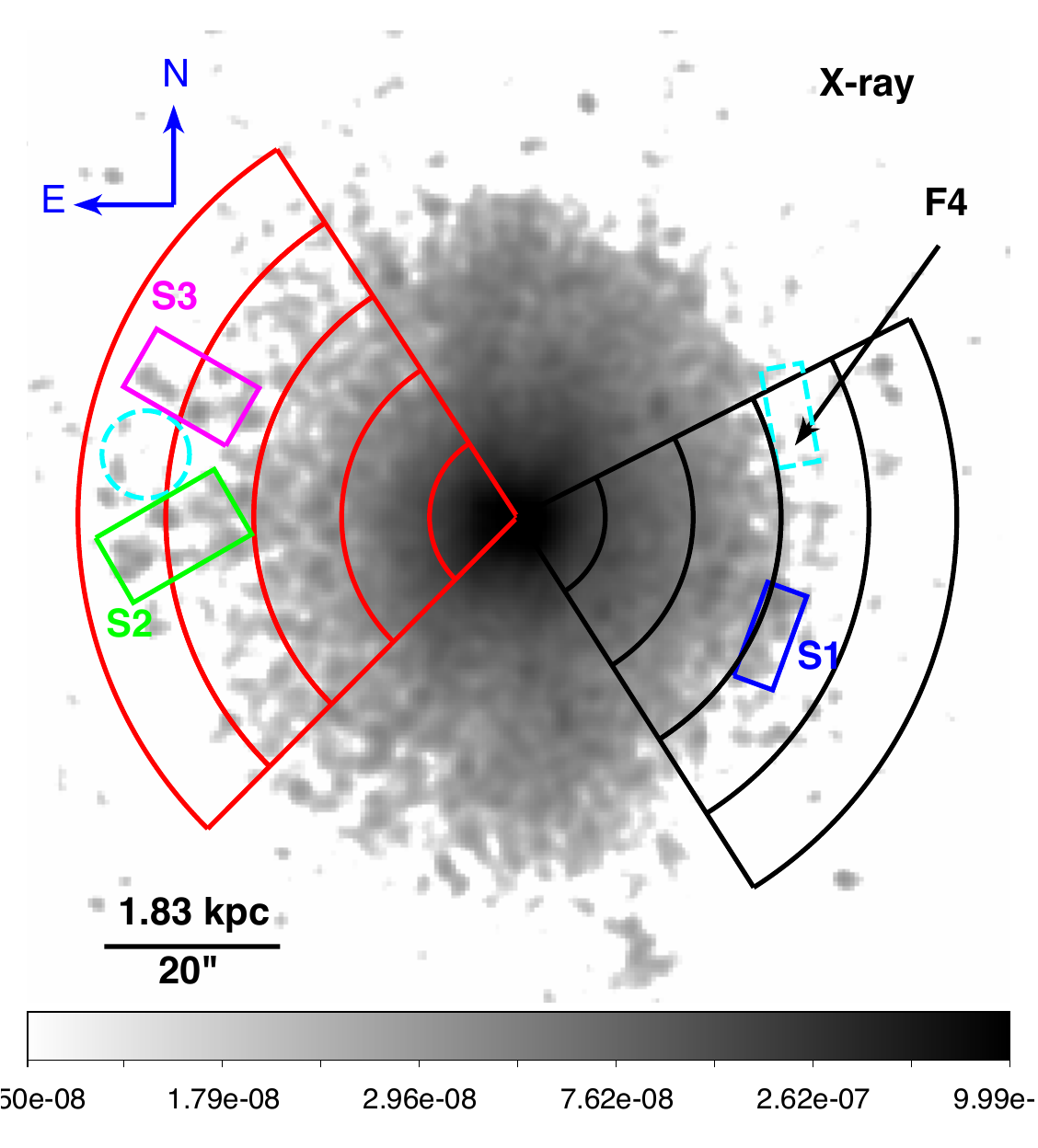} 
      \includegraphics[width=0.45\textwidth]{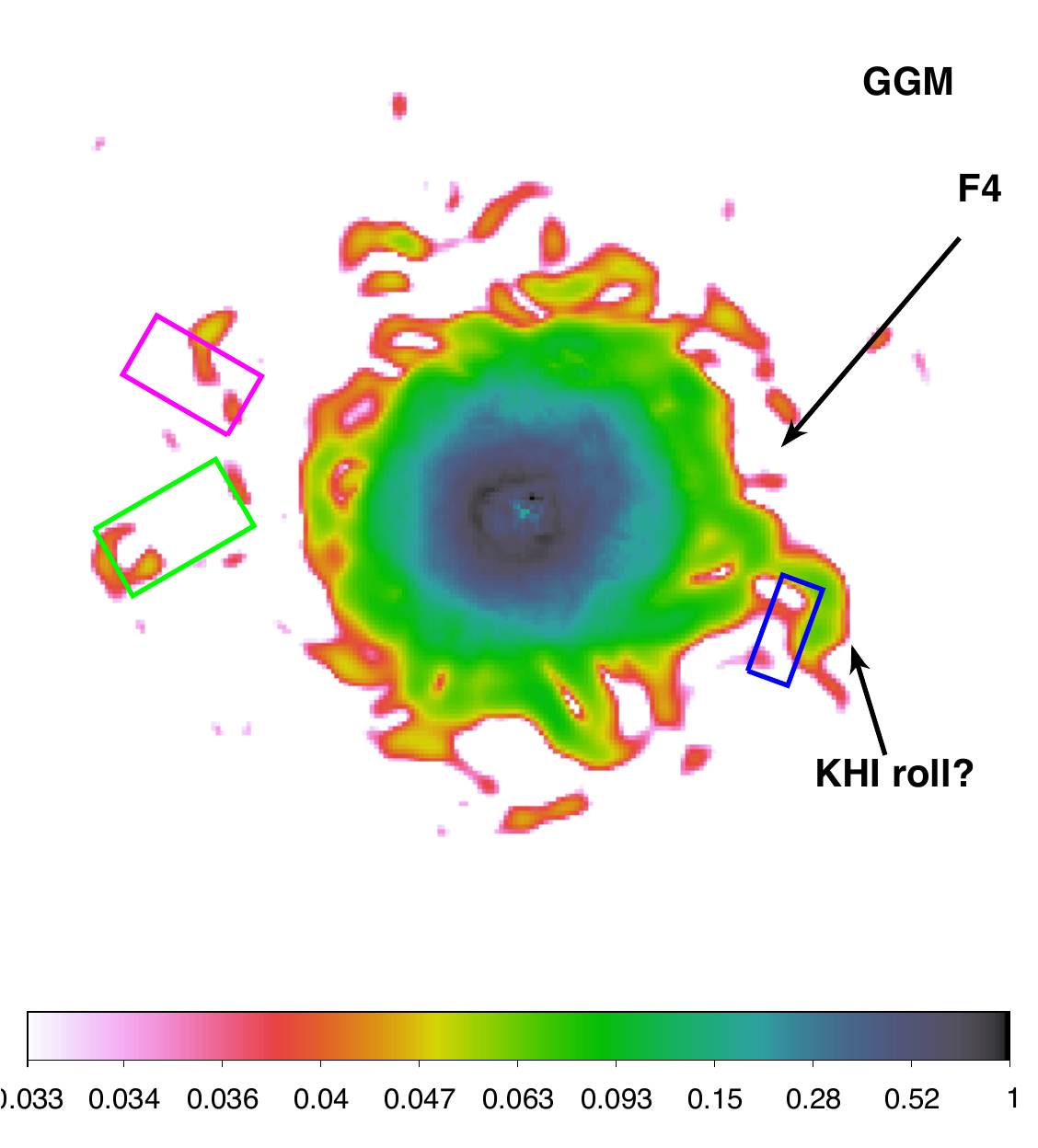} 
                    \includegraphics[width=0.45\textwidth]{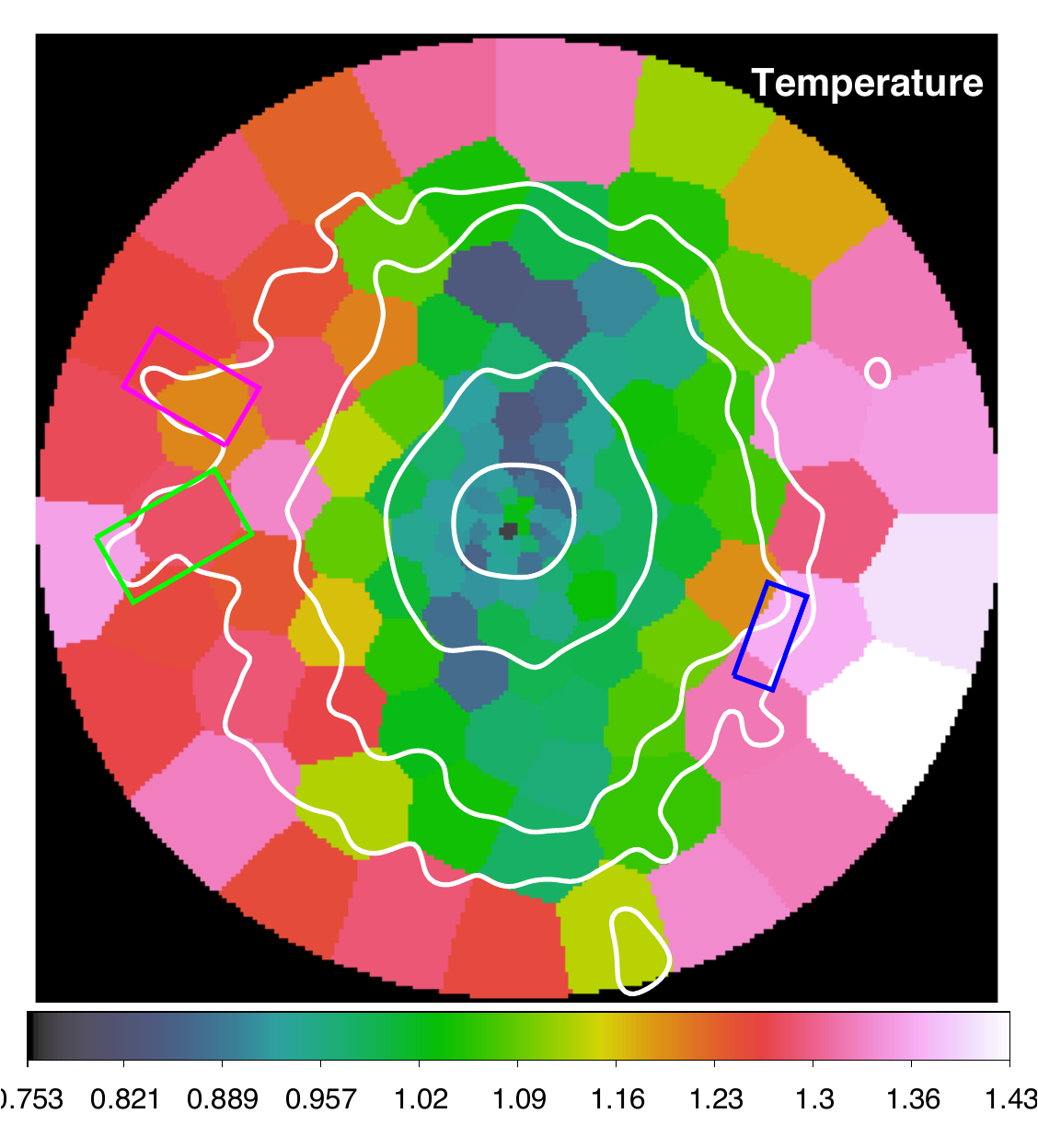} 
                     \includegraphics[width=0.45\textwidth]{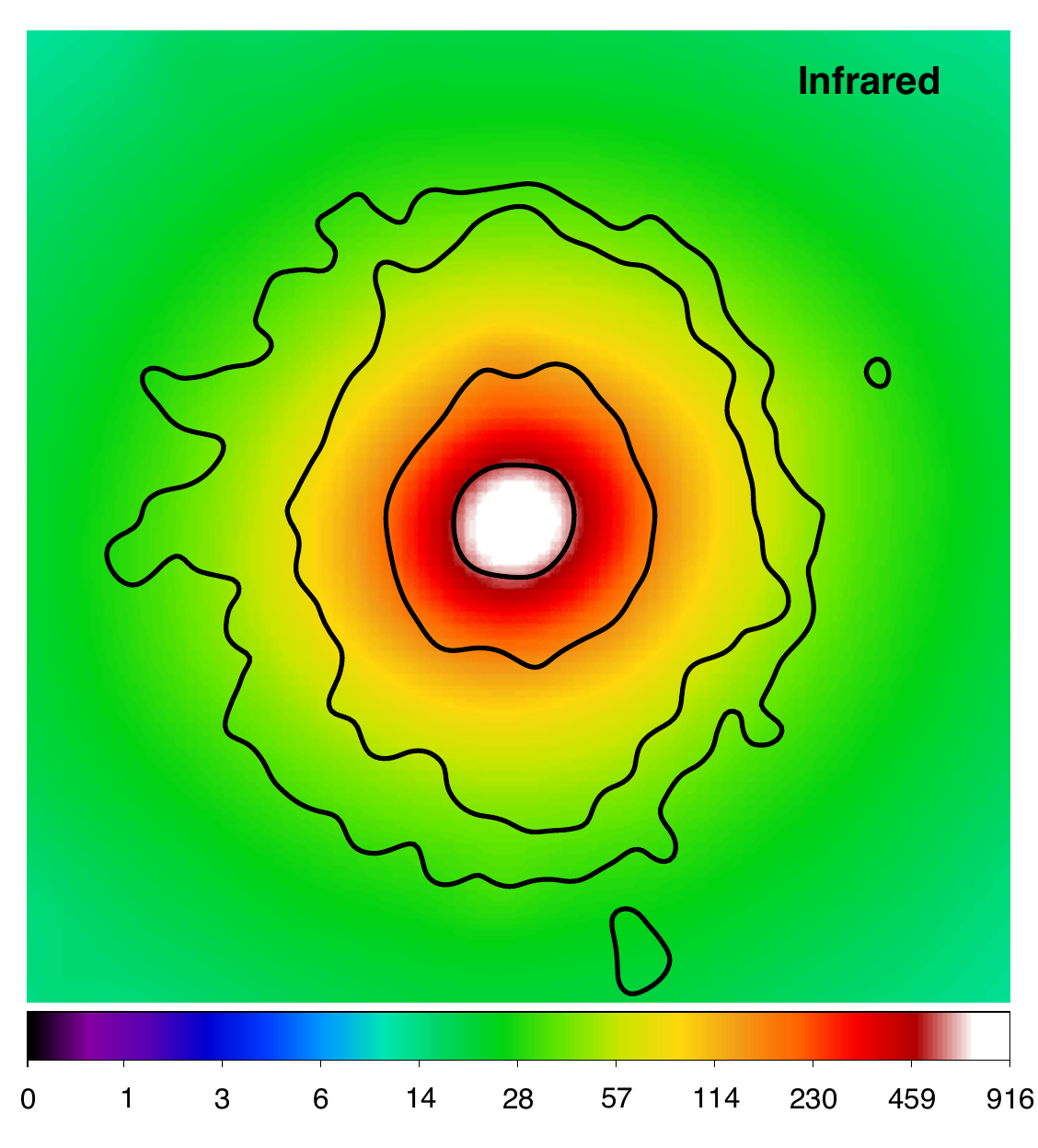} 
\figcaption{\label{fig:img2} Gas properties within the strong cool core of Fornax ($r<4.5$\,kpc; $t_{\rm cool}<1$\,Gyr), corresponding to the yellow dashed circle in Figure~\ref{fig:img}-top-left. {\bf top-left}: {\sl Chandra} X-ray image within a radius of 50$^{\prime\prime}$ (4.5 kpc) with stellar and LMXB components subtracted. Positions of three structures are marked in blue, green, and magenta boxes. Spectra extracted from cyan dashed regions are used as the local background for the spectral fit of S1, S2 and S3. 
{\bf top-right}: Gaussian Gradient Magnitude filtered image of the top-left image, obtained by combining images on scales of 1$\sigma$, 2$\sigma$, 4$\sigma$, 8$\sigma$, and 16$\sigma$. 
{\bf bottom-left}: Two-dimensional temperature distribution of the hot gas in the innermost region of NGC~1399 in units of keV. White contours: {\sl Chandra} X-ray emission in the 0.5--2.0 keV energy band.
{\bf bottom-right}: {\it WISE} infrared image with X-ray contours overlaid. Note that there is no offset between the X-ray and the infrared centroids, implying that Fornax did not experience any recent disturbance.  
}
\end{figure*}

Within its strong cool core ($r<4.5$\,kpc; $t_{\rm cool}<1$\,Gyr), the contribution to the X-ray emissions from the diffuse stellar emission and unresolved 
LMXB may become progressively more important.
We calibrate their contribution to the X-ray surface brightness based on the $K$-band surface brightness profile using the {\sl Two Micron All Sky Survey} ({\sl 2MASS}) (Skrutskie et al.\ 2006) archived image (see Su et al.\ 2017a), which amounts to about 1\% of the total X-ray luminosity. 
We then subtracted the diffuse stellar emission and unresolved LMXB emission from the 0.5--2.0\,keV image as shown in Figure~\ref{fig:img2}-top-left.  
 
\section{\bf analysis and results}

\subsection{Spectroscopic maps}


With {\sl Chandra} observations, we performed a two-dimensional spectroscopic analysis using the Weighted Voronoi Tesselation (WVT) binning (Diehl \& Statler 2006)
based on the Voroni binning algorithm presented in Cappellari \& Copin (2003). We generated a WVT binning image containing 137 regions for the {\sl Chandra} image in the 0.5-2.0 keV band (Figure~\ref{fig:img}-top-left). 
Each bin has a S$/$N of 80.  
We applied a model containing two temperature components to probe its gas metallicity distribution, otherwise ``Fe-bias" would be caused by fitting a single thermal model to multi-phase gas (Buote 2000).  
The two-temperature model takes the form of {\tt phabs}$\times$({\tt vapec}+{\tt vapec}+{\tt pow$_{1.6}$}). The metallicities of the two {\tt vapec} components were linked to each other. The two-temperature fit cannot be well constrained for all regions and we find it necessary to fix one temperature at 1.5\,keV. The other temperature and the normalizations are allowed to vary independently. The resulting Fe abundance map is shown in Figure~\ref{fig:img}-bottom-left. The southwest side of NGC~1399 is more metal-enriched than the northeast side and the metal distribution traces the spiral morphology of the sloshing front.


To probe the thermal structure of the hot gas within its strong cool core ($r<4.5$\,kpc),  
we produced a binned image containing 126 regions for the image in the 0.5-2.0 keV band (Figure~\ref{fig:img2}-top-left). Each bin has a S$/$N of 36. The spectra were fit with the model {\tt phabs}$\times$({\tt apec}+{\tt pow$_{1.6}$}). The abundance was fixed to the solar abundance, which is approximately the average metal abundance at the cluster center. The resulting temperature map is presented in Figure~\ref{fig:img2}-bottom-left.

\subsection{Gas sloshing}

\begin{figure}
   \centering
    \includegraphics[width=0.5\textwidth]{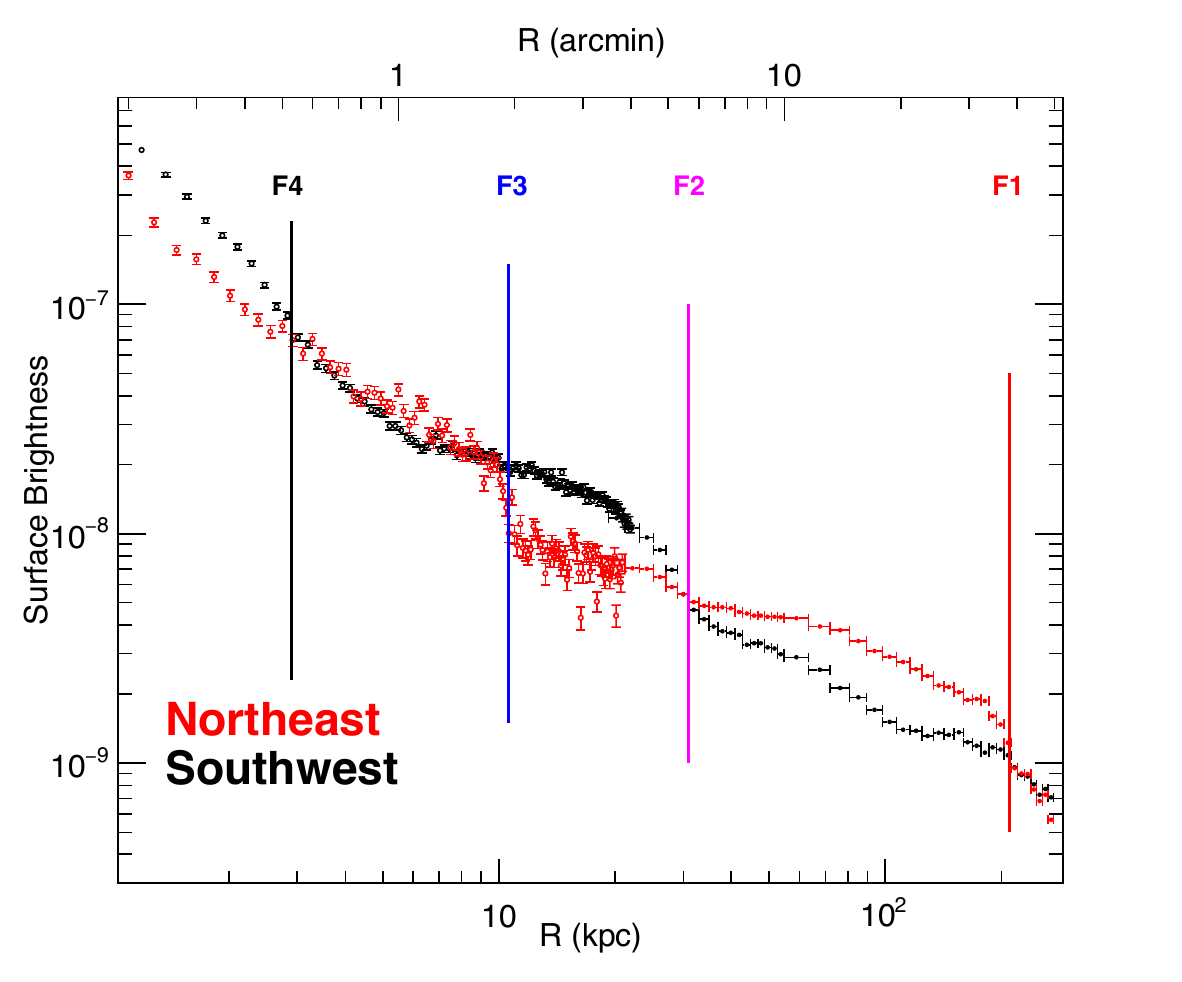}
\figcaption{\label{fig:allsur} The alternating pattern of the surface brightness profiles in the 0.5--2.0 keV energy band of the northeast (red) and the southwest (black) directions over a radial range of 1--250\,kpc. The data points were derived with {\sl Chandra} (small radii, open circles) and {\sl XMM-Newton} (large radii, filled circles) observations and in units of photon\,cm$^{-2}$\,s$^{-1}$\,arcsec$^{-2}$. Vertical solid lines mark the positions of surface brightness edges.}
\end{figure}

Sloshing cold fronts are identified by eye in images and confirmed by abrupt changes in surface brightness and temperature. 
The {\sl XMM-Newton} mosaic image as shown in Figure~\ref{fig:fox}-left reveals several edges in the X-ray surface brightness in Fornax: the outermost one is at 200\,kpc (30$^{\prime}$) to the northeast (F1, red), the second outermost one at 30\,kpc (5$^{\prime}$) to the southwest (F2, magenta), and an inner one at about 10\,kpc (2$^{\prime}$) to the northeast (F3, blue). These fronts stand out in the matching residual image as shown in Figure~\ref{fig:fox}-right, obtained by dividing the {\sl XMM-Newton} image with its azimuthal average. 
Together, they form a spiral pattern wrapping around the bright cluster core.
The deep {\sl Chandra} observation covers F3 and the extended emission to the southwest (Figure~\ref{fig:img}). 
{\sl Chandra} reveals another possible front at $\lesssim3$\,kpc (0.5$^{\prime}$) to the west (F4, black) as shown in Figure~\ref{fig:img2}-top-left.
We present the surface brightness profiles over a radial range of 1--250\,kpc in the northeast and southwest directions (Figure~\ref{fig:allsur}). All four edges can be identified spanning two orders of magnitude in radius. These features are suggestive of gas sloshing triggered by perturbations of minor mergers or off-axis mergers on a large scale.

 \begin{figure}
   \centering
    \includegraphics[width=0.5\textwidth]{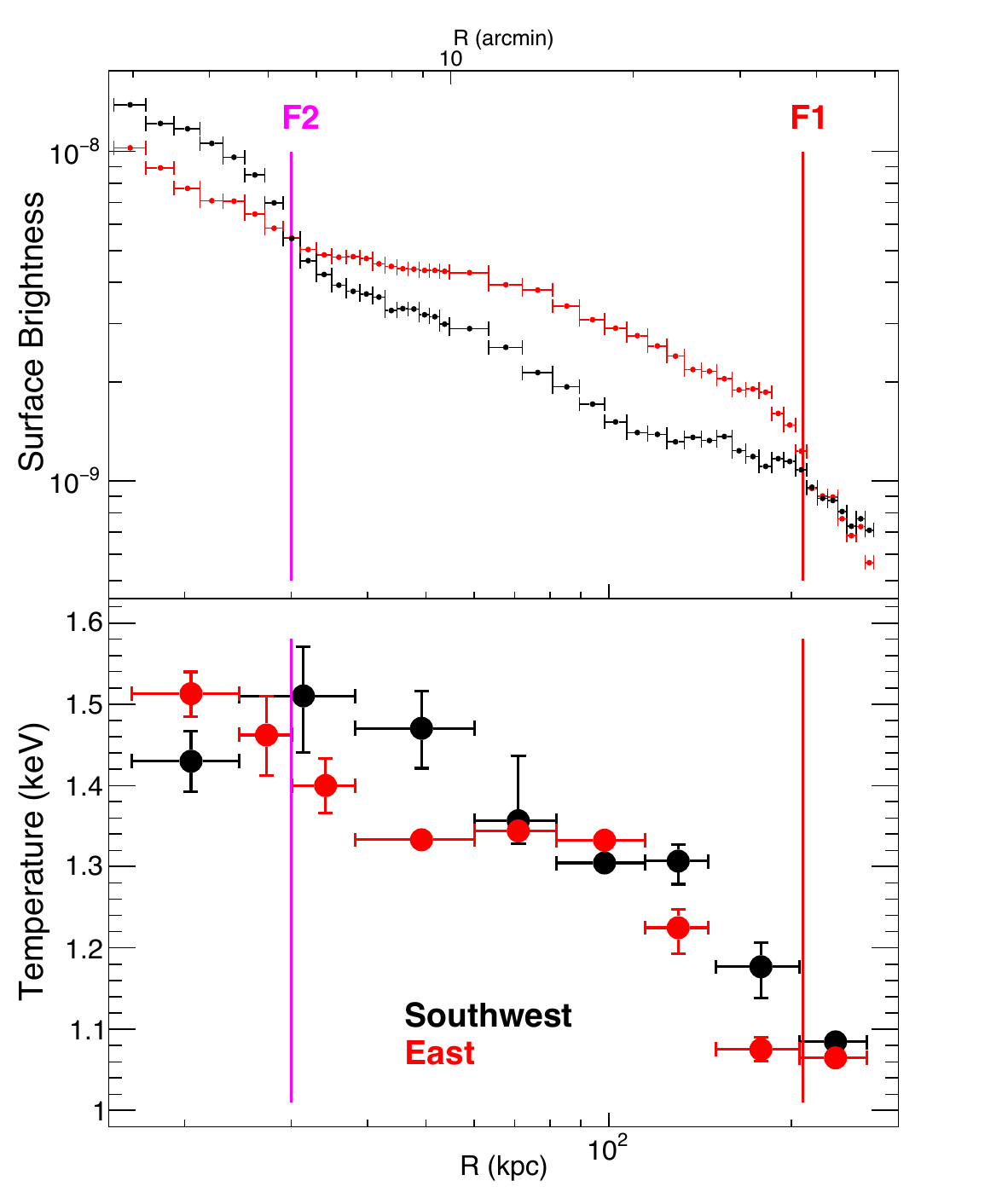}
\figcaption{\label{fig:xmm} {\sl XMM-Newton }surface brightness (in the 0.5--2.0 keV energy band and in units of photon\,cm$^{-2}$\,s$^{-1}$\,arcsec$^{-2}$) (top panel) and projected temperature (bottom panel) profiles of the northeast (red) and the southwest (black) directions outside the cluster center, over a radial range of 3-41$^{\prime}$ (15--250\,kpc). Vertical lines mark the location of the cold fronts F1 and F2.}
\end{figure}

\begin{figure}
   \centering
    \includegraphics[width=0.5\textwidth]{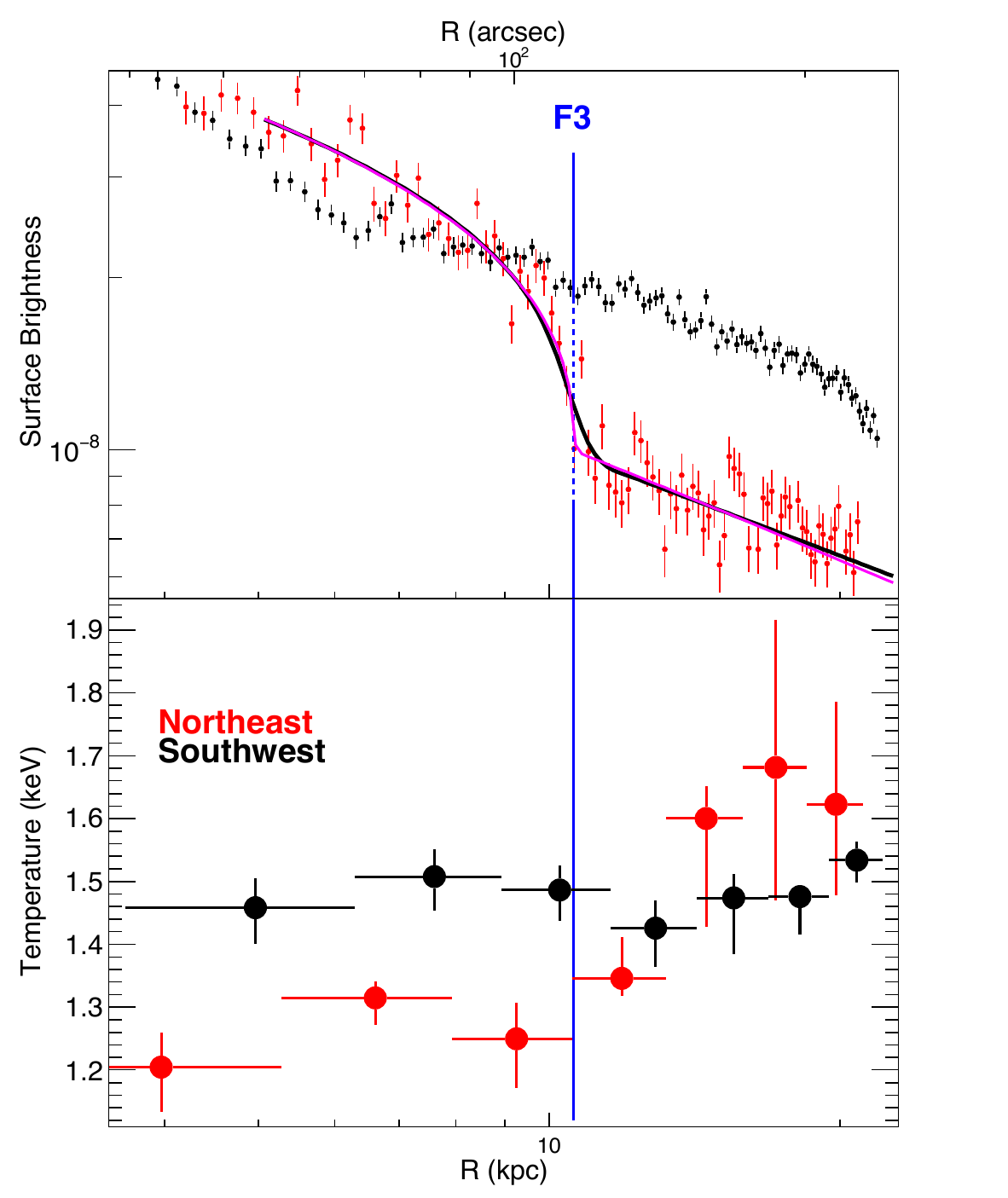}
\figcaption{\label{fig:sb} {\bf top}: {\sl Chandra} surface brightness profiles in the 0.5--2.0 keV energy band 
in the northeast (red) and southwest (black) directions relative to the cluster center and in units of photon\,cm$^{-2}$\,s$^{-1}$\,arcsec$^{-2}$. A sharp edge, marked by the blue line, can be identified in the northeast direction at $R=116^{\prime\prime}$
indicating the presence of a sloshing cold front, while the hot gas is smoother and more extended in the southwest direction. Magenta line: the best-fit broken power-law density model corresponding to a density jump of 2.1. Black line: best-fit broken power-law density model smeared with the Gaussian $\sigma=5^{\prime\prime}$. {\bf bottom}: Projected temperature profiles in the northeast and southwest directions, respectively (red and black sectors in Figure~\ref{fig:img}-top-right) measured with {\sl Chandra}. Radius is relative to the cluster center.} 
\end{figure}

We present the temperature profiles of the east and southwest directions crossing the outer fronts in Figure~\ref{fig:xmm}. The two edges, F1 and F2, are associated with cooler gas relative to the gas at the same radii but on the opposite side of the cluster. 
The northeast front at $r=10.6$\,kpc (
$116^{\prime\prime}$), F3, is the most evident front (Figure~\ref{fig:img}-top-left). 
We apply a Gaussian Gradient Magnitude (GGM) filter to highlight sharp edges in the {\sl Chandra} X-ray image, which determines the magnitude of surface brightness gradients using Gaussian derivatives (Sanders et al.\ 2016b; Walker et al.\ 2016). Brighter regions correspond to sharp features in surface brightness. The resulting GGM image is shown in Figure~\ref{fig:img}-bottom-right, obtained by combining images on scales of 2$\sigma$, 4$\sigma$, 8$\sigma$, 16$\sigma$, and 32$\sigma$. F4 stands out as a sharp edge at $r=10$\,kpc to the northeast on the GGM image. 
We compare the {\sl Chandra} X-ray surface brightnesses derived in annular sectors with a radial bin width of $2^{\prime\prime}$ from the cluster center to the northeast (133$^{\circ}$--159$^{\circ}$, marked in the red sector in Figure~\ref{fig:img}-top-right) and to the southwest (290$^{\circ}$--390$^{\circ}$, marked in the black sector in Figure~\ref{fig:img}-top-right). As shown in Figure~\ref{fig:sb}, the surface brightness profile of the southwest sector declines more smoothly than that of the northeast sector. 
We fit a broken power-law density
model to the northeast profile and we obtain a break at $116\pm1^{\prime\prime}$ (10.6\,kpc) relative to the cluster center, which corresponds to a gas density jump of $2.1\pm0.2$.
We convolve this power-law density model with a Gaussian component and obtain a best-fit width of $\sigma=5^{\prime\prime}$\,(450\,pc). The smoothed model does {\it not} provide a better fit with a $F$-test probability of 0.11.
We extracted spectra from seven concentric annular sectors across this northeast edge (marked in the red annular sectors in Figure~\ref{fig:img}-top-right). The spectra were fit to a single-temperature model. For comparison, we performed the same analysis for the southwest direction (marked in the black annular sectors in Figure~\ref{fig:img}-top-right). 
The resulting temperature profiles are presented in Figure~\ref{fig:sb}. 
The gas on the brighter side of the northeast edge is cooler than that on the fainter side, suggesting that this edge is a cold front. 
The Fe abundance is higher within the edge than that outside as shown in Figure~\ref{fig:img}-bottom-left. 
Hot, diffuse, and relatively pristine cluster gas from the larger radii may have been brought into contact with the cool, dense, and enriched gas by sloshing. 
The distribution of the gas to the southwest over the same radial range displays a relatively uniform thermal and chemical distribution.

The X-ray emitting structure within the strong cool core of Fornax, tracing the pure hot gas distribution, is shown in Figure~\ref{fig:img2}-top-left. We note a surface brightness discontinuity to the west with a best-fit edge at $r=2.9$\,kpc (Figure~\ref{fig:inn}-top), while the surface brightness profile to the east is relatively smooth. 
We derive the temperature profile across this west edge as shown in Figure~\ref{fig:inn}-bottom. It rises abruptly from $\sim1$\,keV to $\sim1.5$\,keV over a radial range of 1--2 kpc, manifesting the presence of a cold front, consistent with the temperature map (Figure~\ref{fig:img2}-bottom-left). In contrast, the temperature profile over the same radial range to the east varies from $\sim1.1$\,keV to $\sim1.2$\,keV. 

\subsection{Sub-kpc structures}

\begin{figure}
\vspace{0.3cm}
   \centering
       \includegraphics[width=0.5\textwidth]{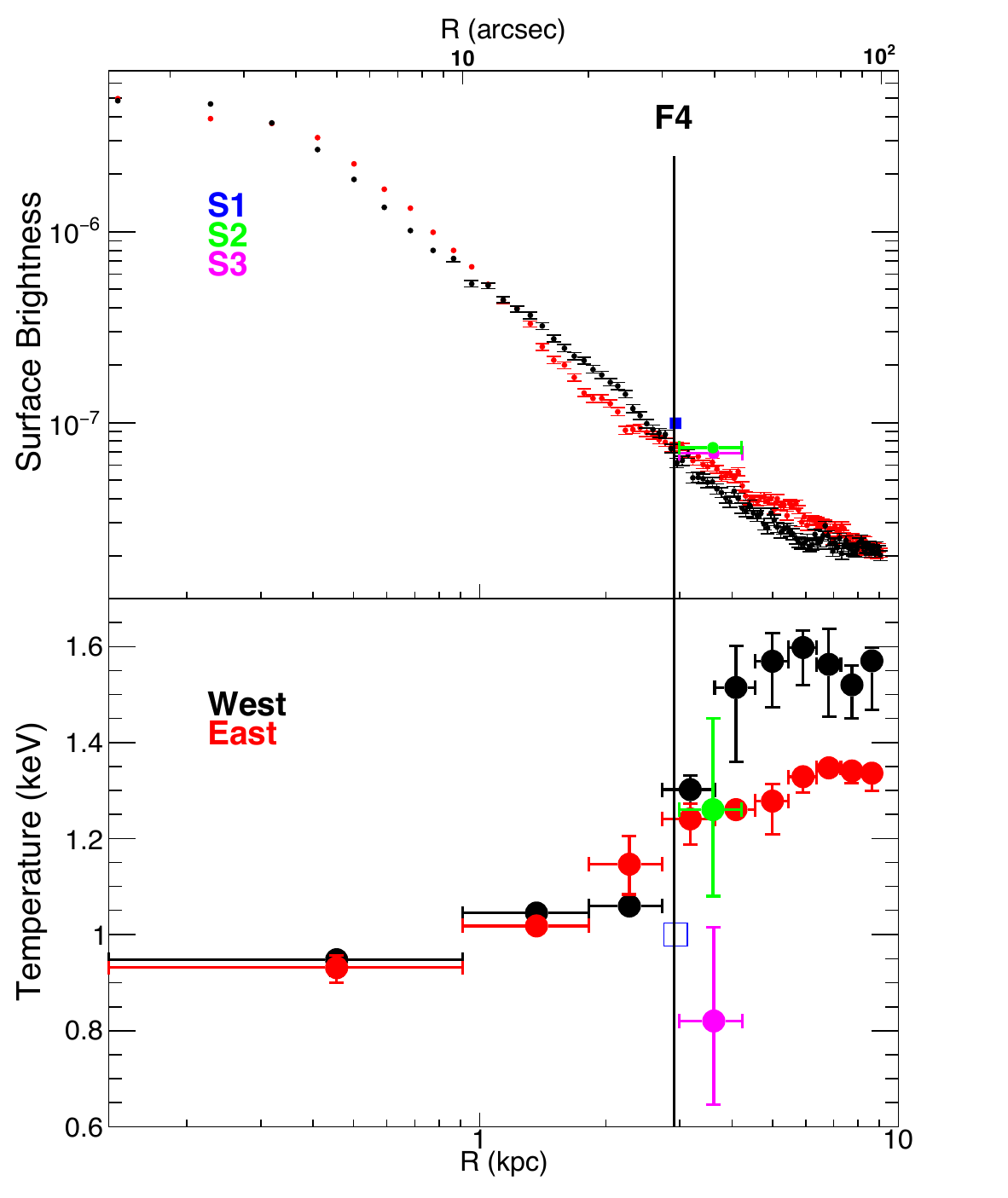} 
\figcaption{\label{fig:inn} {\sl Chandra} surface brightness (in the 0.5--2.0 keV energy band and in units of photon\,cm$^{-2}$\,s$^{-1}$\,arcsec$^{-2}$) (top panel) and projected temperature (bottom panel) profiles of the east (red) and the west (black) directions within a radius of 100$^{\prime\prime}$ (9.1 kpc). The corresponding regions of the inner five data points on the temperature profile are marked in Figure~\ref{fig:img2}-top-left. The location of a possible sloshing cold front is marked by the black line. The three structures (Figure~\ref{fig:img2}) are marked in blue (S1), green (S2), and magenta (S3) symbols; their temperatures are derived with local background. The temperature of S1 is not constrained in the spectral fit. We derive its temperature (blue open square) by assuming S1 is  in pressure equilibrium with the ambient ICM.}
\end{figure}


We produce a GGM image to probe substructures in the innermost 5\,kpc by 
combining GGM images on scales of 1$\sigma$, 2$\sigma$, 4$\sigma$, 8$\sigma$, and 16$\sigma$. 
As shown in shown in Figure~\ref{fig:img2}-top-right, a sub-kpc region of enhanced surface brightness is visible just outside the cold front (F4) to the west (blue box). Two 3--5\,kpc long bright filaments can be identified to the east at a similar radius (Figure~\ref{fig:img2} green and magenta boxes). We perform spectral analysis for these structures to determine their natures. We apply local background spectra extracted from neighboring regions as shown in Figure~\ref{fig:img2}-top-left (cyan dashed box for S1; cyan dashed circle for S2 and S3). The spectra were fit to the model {\tt phabs}$\times$({\tt apec}+{\tt pow$_{1.6}$}). Their metal abundance was fixed {at} the solar abundance. 
The best-fit temperatures of S2 and S3 are $1.26^{+0.19}_{-0.18}$\,keV and $0.82^{+0.19}_{-0.18}$\,keV, respectively (Figure~\ref{fig:inn}-bottom). Their ambient ICM has a temperature of $\sim1.3$\,keV. 
{To calculate their densities, we assume these structures are cylinders in 3D with a volume $V=l\cdot\pi(w/2)^2$, which is a common approximation for elongated substructures at cluster centers (e.g., David et al.\ 2017).} 
We derive the pressure of the ambient ICM using the deprojected density profile of NGC~1399 determined in Paper 1. 
S3 is in pressure equilibrium with the ambient ICM. S2 would be over-pressurized unless its temperature is near its lower limit of $\sim1$\,keV.  
The temperature of S1 cannot be constrained. Assuming S1 is also in pressure equilibrium, its temperature would be $\sim1$\,keV.
The low temperature (cooler than the ambient ICM) of these sub-kpc features meets our expectation for KHI eddies growing at cold fronts (e.g., NGC~1404, Su et al.\ 2017a). Their properties and locations are also consistent with
being low entropy filaments induced by thermal instability (e.g., NGC~5044, David et al.\ 2014; 2017).

\section{\bf Discussion}

We perform a joint {\sl Chandra} and {\sl XMM-Newton} analysis of the Fornax Cluster.
A series of sloshing cold fronts (F1, F2, F3, and F4) are present in the ICM. Nearly all of them fall on the same spiral structure, spanning from a radius of 200\,kpc to the cooling radius.   
We discuss their constraints on the cluster merging history and the effect of gas sloshing in the chemical and thermal distribution of the ICM.

\subsection{Merging history recorded by sloshing cold fronts}

\begin{figure}
   \centering
    \includegraphics[width=0.5\textwidth]{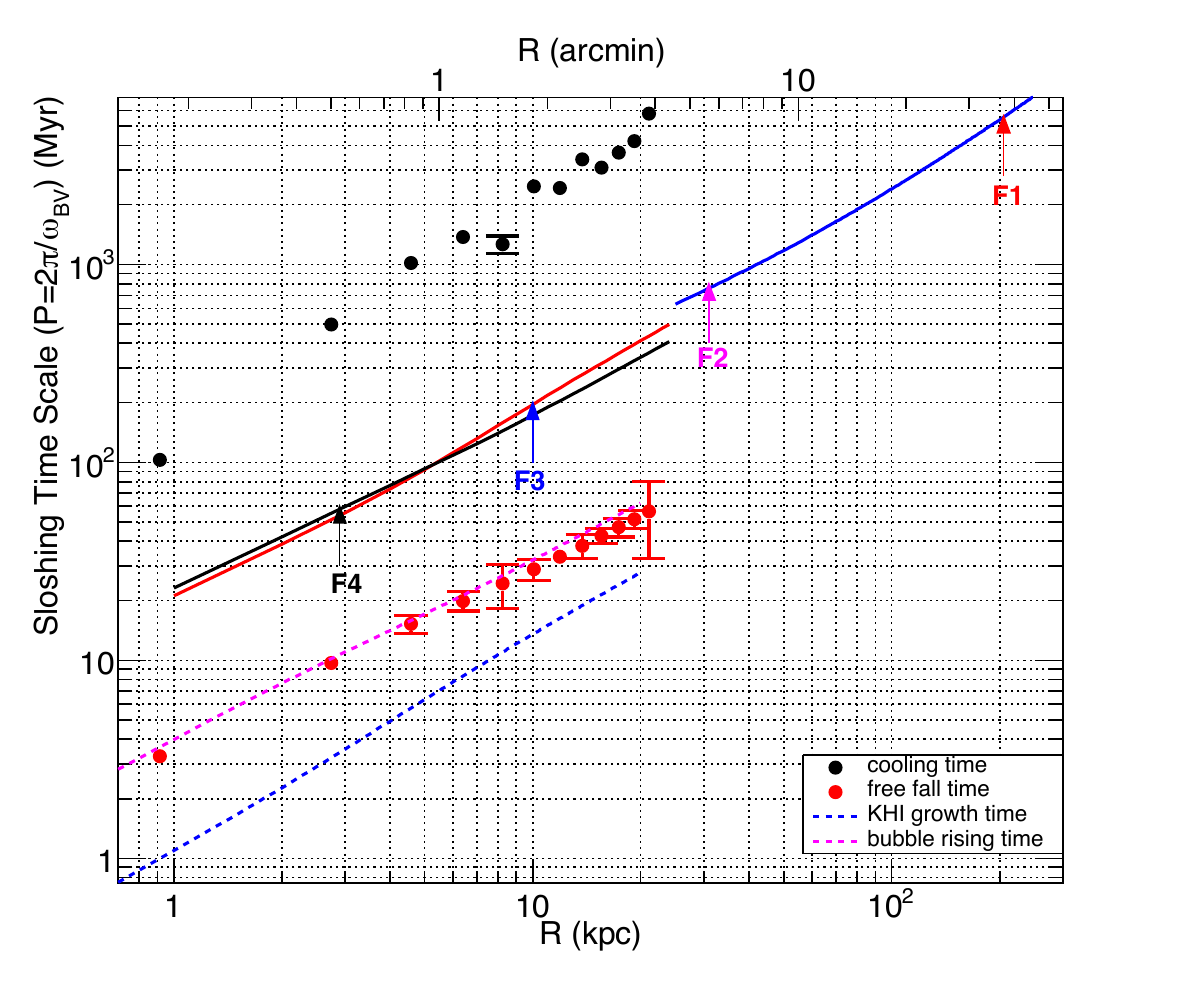}
\figcaption{\label{fig:sloshing} The sloshing time scale (oscillation period, $P=2\pi/\omega_{\rm BV}$, where $\omega_{\rm bv}$ is the Brunt-V$\ddot{\rm a}$s$\ddot{\rm a}$l$\ddot{\rm a}$ frequency) as a function of radius, compared to the cooling time (black circles) and the free fall time (red circles). Data points within $r=20$\,kpc are obtained with {\sl Chandra} and those outside $r=30$\,kpc are with {\sl XMM-Newton}. Dark black, red, and blue solid lines are derived using hydrostatic mass, velocity dispersion mass, and the best-fit NFW mass, respectively. Arrows mark the positions of the four sloshing fronts found in Fornax.}
\end{figure}

We identify four edges in the X-ray surface brightness at radii of 
200\,kpc, 30\,kpc (Figures~\ref{fig:fox} and \ref{fig:xmm}), 10\,kpc (Figures~\ref{fig:img} and \ref{fig:sb}), and 3\,kpc (Figures~\ref{fig:img2} and \ref{fig:inn}), all distributed along the SW-NE direction (Figures~\ref{fig:fox} and \ref{fig:allsur}). 
The brighter sides of these edges comprise cooler gas. 
We infer that they are sloshing cold fronts induced by minor or offset mergers.
Gas motion around the cluster center approximates the oscillating flow in a statically stable environment.
We calculate the Brunt-V$\ddot{\rm a}$s$\ddot{\rm a}$l$\ddot{\rm a}$ frequency (buoyancy frequency; Cox 1980) at each radius, $r$ (Balbus \& Soker 1990):
\begin{equation}
\omega_{\rm BV}=\Omega_{\rm K}\sqrt{\frac{1}{\gamma}\frac{d {\rm ln} K}{d {\rm ln} r}},
\end{equation}
where $K=kT/n^{2/3}$ is the gas entropy, $\Omega_{\rm K}=\sqrt{\frac{GM}{r^3}}$ the Keplerian frequency, and $\gamma=5/3$.
The density and temperature profiles of the cluster center, measured with {\sl Chandra} observations, are used to calculate $K$ and to derive the hydrostatic mass profile, $M(r)$; for comparison, we also derive the total mass using the stellar velocity dispersion (Paper 1). Gas properties outside the weak cool core are measured with {\sl XMM-Newton}; the best-fit Navarro-Frenk-White (NFW) dark matter profile of Navarro et al.\ (1997) is used to calculate $\Omega_{\rm K}$.
The resulting sloshing period of $P=2\pi/\omega_{\rm BV}$ is shown in Figure~\ref{fig:sloshing}, which increases with radius. 
{The sloshing timescale ($P$ or $P/2$) provides an order-of-magnitude estimate of the age of the sloshing front (Churazov et al.\ 2003). To verify this approximation, we compare these time scales to the evolution of the sloshing simulated for a Virgo-like cluster (Figure~7 in Roediger et al.\ 2011). The ages of the simulated sloshing fronts at radii of 100\,kpc (NW), 40\,kpc (SE), and 20\,kpc (NW) are 1.7\,Gyr, 0.8\,Gyr, and 0.6\,Gyr, respectively (fronts within 10\,kpc are not resolved). These scales are comparable to the sloshing fronts observed in Fornax in this work at 200\,kpc (E), 30\,kpc (SW), and 10\,kpc (NE) with their ages ($P/2\sim P$) corresponding to 2.5--5.0\,Gyr, 0.4--0.7\,Gyr, 0.1--0.2\,Gyr, respectively, as shown in Figure~\ref{fig:sloshing}. Our approach thus leads to reasonable approximations. In future work, we plan to refine our estimates with simulations specifically tailored to Fornax (Sheardown et al.\ in preparation).}

Simulations reveal that the infall of one subcluster is capable of inducing multiple sloshing cold fronts as the displaced gas peak oscillates back and forth around the dark matter center of the main cluster (ZuHone et al.\ 2010; 2011). 
That F1, F2, and F3 fall on the same spiral structure is typical for sloshing cold fronts produced in a single merger event, as seen in simulations (e.g., Roediger et al.\ 2011).
NGC~1404 and NGC~1387 are the second and third brightest member galaxies in Fornax (Figure~\ref{fig:fox}). NGC~1387 is more than $5\times$ fainter in X-rays than NGC~1404 and its radial velocity relative to NGC~1399 is $\gtrsim100$\,km\,s$^{-1}$, $5\times$ smaller than that of NGC~1404. Therefore, the infall of NGC~1404 may have initiated the gas sloshing.
Adopting a time scale of $P/2\sim P=2.5\sim5$\,Gyr for the outermost front at $r=200$\,kpc (Figure~\ref{fig:sloshing}), we estimate that the infall of NGC~1404, took place more than 2\,Gyr ago, longer than the typical crossing time of galaxy clusters. This is consistent with our previous study of NGC~1404 approaching the inner region of Fornax for the second time (Su et al.\ 2017b). Fornax is one of the few clusters where the perturber that has initiated the sloshing can be identified, providing an ideal clean case for tailored simulations.

\subsection{Metallicity redistribution at the cluster center}

The two-dimensional Fe abundance distribution of NGC~1399 (Figure~\ref{fig:img}-bottom-left) shows that the hot gas with enhanced metallicity ($>1.0$\,Z$_{\odot}$) reaches radii of approximately 5--10\,kpc to the north and the east and it 
is more extended to the south and the west, reaching beyond 12\,kpc. The extended distribution of enriched ICM may reflect an
AGN outburst or gas sloshing.
The enriched gas to the north is most likely due to the AGN outburst along the north-south direction.  
Kirkpatrick \& McNamara (2015) calibrated an empirical relation between the maximum projected radius of the uplifted gas and the cavity enthalpy 
\begin{equation}
R_{\rm Fe}=(57\pm30)\times {H}^{0.33\pm0.08}\,{\rm kpc},
\end{equation}
where $H$ is in units of $10^{59}$ ergs.
Substituting the cavity enthalpy of NGC~1399 (Paper~1),  we obtain a $R_{\rm Fe}$ of 3--10\,kpc, consistent with the observed enriched gas extent to the north.  
The enriched gas extended from the south to the west follows the spiral pattern of the gas sloshing, strongly suggesting that gas sloshing is driving the metal redistribution at the center cluster.     

While AGN bubbles are very vulnerable to gas motions at the cluster center (Morsony et al.\ 2010), the two bubbles in NGC~1399 are remarkably symmetric and intact in both X-ray and radio. We thus infer that the AGN outburst has taken place later than the gas sloshing. This is consistent with the cycle of AGN outbursts (a few tens of Myr) being much shorter than the time scale of gas sloshing ($\sim1$\,Gyr) (ZuHone et al.\ 2010). 
The enriched gas along the southwest front is more metal abundant, more extended, and distributed to a larger radius than the enriched gas to the north. We therefore conclude that gas sloshing, rather than AGN uplift, has played a primary role in redistributing the enriched gas, at least in this particular case. 

\subsection{Microphysics and regulation of cooling}

While the cooling time drops below 1\,Gyr at the center of Fornax, as in
many more massive cool-core clusters, the star formation rate is negligible in NGC~1399 (Vaddi et al.\ 2016). {AGN feedback is by far the most viable solution to the cooling problem.
The mechanical energy provided by AGN is correlated with the cool-core luminosity (B{\^i}rzan et al.\ 2004). AGN activity can respond to cooling through precipitation of cooled gas, as proposed by McCourt et al.\ (2012). 
Still, it is unclear how the jet power is transformed into thermal energy. One route is by the dissipation in the ICM of turbulence generated in the wakes of rising bubbles (Churazov et al.\ 2002).
Then again, Zhuravleva et al.\ (2017) found that in the cool cores of some clusters (e.g., Perseus and Abell~1795), gas perturbations are associated with gas motion (isobaric) rather than AGN bubbles (isothermal), suggesting that the bulk of the turbulence may not be generated by radio outbursts.} 
Apart from AGN outbursts, gas sloshing has been considered as a means of quenching cooling flows (Fujita et al.\ 2004; ZuHone et al.\ 2010). 
Frequent gas sloshings do not necessarily require a high merging rate in that multiple sloshing fronts can be induced by just one subcluster infall (ZuHone et al.\ 2010; 2011). 
NGC~1399 displays a series of sloshing cold fronts at various radii. 
As demonstrated in Figure~\ref{fig:sloshing}, the sloshing time scale is $\approx10\times$ shorter than the cooling time over the entire cluster center, making it a promising mechanism to heat the cool core.
However, it would be far fetched for gas sloshing to be the primary mechanism of preventing the gas from cooling,
unless the frequency of gas sloshing responds to the state of the gas. 
{Nevertheless, with AGN feedback being the primary solution, gas sloshing may still provide supplemental heating to the cool core,
as long as heat can be transported effectively between gases of different entropy.}
Below we discuss two transport processes that may occur at the interface.

\subsubsection{Conduction}

Thermal conduction is expected to erase temperature gradients outside the strong cool core ($1\,{\rm Gyr}<t_{\rm cool}<14$\,Gyr) (Voit et al.\ 2015), where most sloshing cold fronts reside.
If conduction is reduced, low entropy gas brought out by sloshing would eventually fall back to the cluster center (Ghizzardi et al.\ 2014).

We calculate the characteristic mean-free-path (mfp) of electrons, $\lambda_e$, in the hot plasma at the leading edge. Sarazin (1988) gives:
\begin{equation}
\lambda_e=\frac{3^{3/2} (kT_e)^2}{4\pi^{1/2}n_e{q_e}^4\rm ln \Lambda}.
\end{equation}
where $n_e$ is the electron density and the Coulomb logarithm is \begin{equation}{\rm ln} \Lambda = 35.7+ {\rm ln} \left[\left(\frac{kT_e}{\rm 1\,keV}\right)\left(\frac{n_e}{\rm 10^{-3} cm^{-3}}\right)^{-1/2}\right].\end{equation}
Equation (3) can be approximated by
\begin{equation}
\lambda_e\approx340\,{\rm pc}\left(\frac{kT_e}{{\rm 1\,keV}}\right)^2\left(\frac{n_e}{10^{-3}\,{\rm cm}^{-3}}\right)^{-1},
\end{equation}
assuming the temperatures of electrons, ions, and gas are equal. Substituting a temperature of 1.4\,keV and a gas density of $0.004$\,cm$^{-3}$ for the northeast cold front at $r=10$\,kpc (F3, the most prominent cold front), we obtain a $\lambda_e$ of 150\,pc. 
This value is smaller than we can resolve in either our imaging or
spectral analysis. 
Therefore, we are unable to strictly determine the effective conductivity in the ICM in this current work.
Following Sarazin (1998) and Markevitch et al.\ (2003), we estimate the time scale for the thermal conduction to operate in the Spitzer regime
\footnotesize
\begin{equation}
t_{\rm cond}\sim kl^2n_{\rm e}/\kappa_{\rm s}
\sim 12\left(\frac{n_e}{0.002\,{\rm cm^{-3}}}\right)\left(\frac{l}{100\,{\rm kpc}}\right)^2\left(\frac{T}{10\,{\rm keV}}\right)^{-5/2}\,{\rm Myr},
\end{equation}
\normalsize
where $\kappa_{\rm s}$ is the Spitzer value (1956), $l$ the size of the cold front, and $T$ the ICM temperature. Applying the conditions of F3, we obtain $t_{\rm cond}\lesssim10$\,Myr, shorter than the age of the bubbles ($\sim15$\,Myr, Paper 1). As we discussed in \S4.2, AGN outburst is more recent than the event that caused the sloshing. 
The northeast sloshing front, residing at 10\,kpc, has a sloshing time scale of $100\sim200$\,Myr (Figure~\ref{fig:sloshing}). 
The conductivity, possibly being reduced by a factor of $(\kappa/\kappa_{\rm s})^{-1}\sim t_{\rm age}/t_{\rm cond}\gtrsim$\,10, is not sufficient to wipe out temperature gradient. 
This is consistent with the edge-fitting of the surface brightness profile that convolving the power-law density model with a Gaussian component is not required (\S3.2). 
Komarov et al.\ (2014) argue that fluid elements along the presumably-incompressible cold front are stretched, which tend to enhance the temperature gradients and align the originally-random magnetic field lines along the front. Consequentially, and counterintuitively, heat flux can be reduced at the cold front where the temperature gradient is the largest. {Then again, 
using MHD simulations, ZuHone et al.\ (2013, 2015) found that the temperature gradient can nevertheless be reduced if the conduction is anisotropic even in the presence of this magnetic layer.}

\subsubsection{KHI and turbulent mixing}
The innermost sloshing cold front (F4) resides at $r\lesssim3$\,kpc to the west while the two AGN bubbles are more than 5\,kpc away from the cluster center. Sloshing brings high entropy gas to the innermost regions where cooling is most severe ($t_{\rm cool}<$\,1Gyr). 
This innermost sloshing cold front is not sharp with a sub-kpc bright structure, a KHI eddy candidate, visible at the interface (blue box in Figure~\ref{fig:img2}). Its temperature could be $\approx1$\,keV, similar to the gas temperature inside the interface.

\begin{figure}
\vspace{0.3cm}
   \centering
       \includegraphics[width=0.5\textwidth]{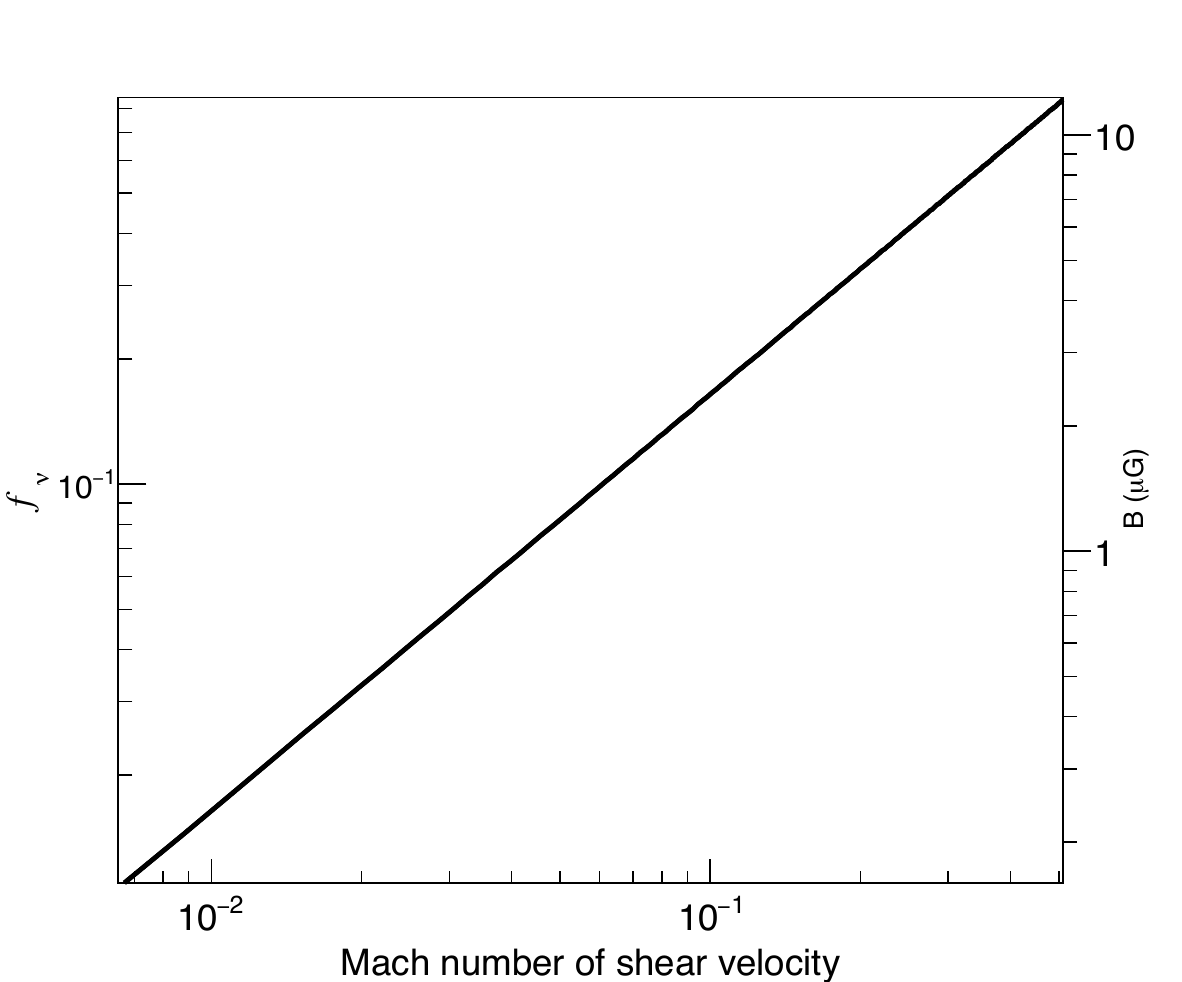} 
\figcaption{\label{fig:khi} For a given shear velocity along the sloshing front (x-axis), the presence of KHI at the innermost sloshing front in NGC~1399 would be allowed if $f_{\nu}$ (left-axis, the fraction of the viscosity relative to the Spitzer value) and $B$ (right-axis, the strength of the ordered magnetic field) stay below the black solid line.}
\end{figure}

Either viscosity or ordered magnetic field can damp out the growth of KHI. 
The presence of a KHI roll allows us to put upper limits on the viscosity and the magnetic field strength using the criteria presented in Roediger et al.\ (2013) and Vikhlinin et al.\ (2001), respectively.
For a shear flow velocity with a Mach number $\mathcal{M}$,
the fraction of the viscosity relative to the Spitzer value, $f_{\nu}$, needs to be smaller than

\footnotesize
\begin{equation}
{f_{\nu}}<\frac{10}{16\sqrt{\Delta}}\cdot\frac{\mathcal{M}c_s}{400\,\rm km/s}\cdot\frac{\ell}{10\,\rm kpc}\cdot\frac{n_{e,h}}{10^{-3}\rm cm^{-3}}\left(\frac{kT_{h}}{\rm 2.4 \,keV}\right)^{-5/2}\hspace{-2.2em},
\end{equation}
\normalsize
where $c_s(=\sqrt{{\gamma kT_h}/{\mu m_p}})$ is the sound speed, $T_h$ and $n_{e,h}$ are the temperature and density of the ICM on the high entropy side, $\ell$ is a characteristic length scale, and $\Delta = \frac{(\rho_{h}+\rho_{c})^2}{\rho_{h}\rho_{c}}$ characterizes the contrast of the gas densities on each side of the interface. 
This KHI eddy candidate has a height of $h\sim0.3$\,kpc; we estimate $\ell\approx2h\sim0.6$\,kpc (Roediger et al.\ 2013). 
Likewise, the strength of an ordered magnetic field needs to be smaller than 

\begin{equation}
B < \left[8\pi\frac{\gamma T_{h}n_{e,h} }{1+T_c/T_h}\right]^{\frac{1}{2}}{\mathcal{M}}.
\end{equation}

\begin{figure}
\vspace{0.3cm}
   \centering
       \includegraphics[width=0.5\textwidth]{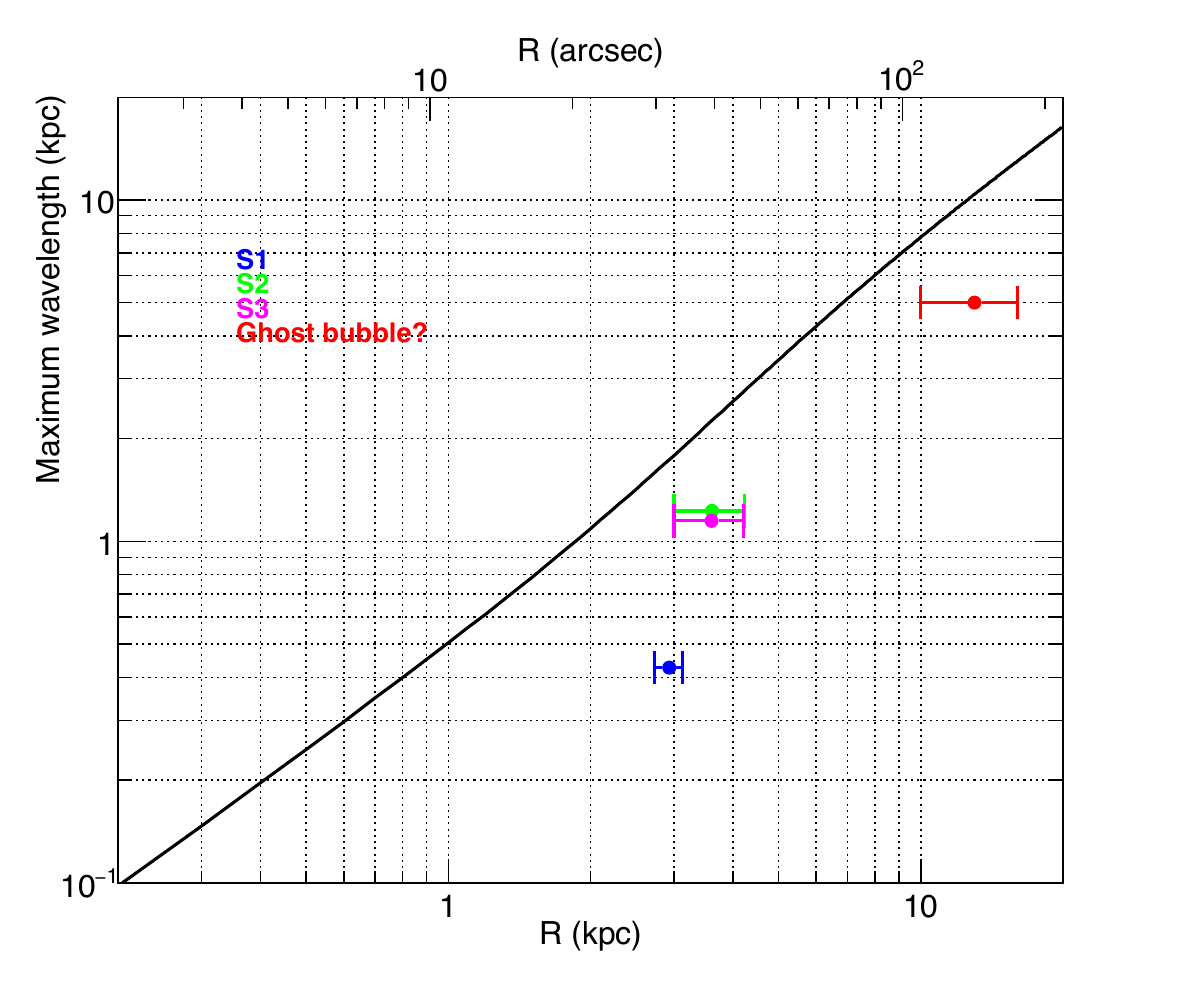} 
\figcaption{\label{fig:khi2} Maximum wavelength for KHI rolls as a function of radius in NGC~1399 (black solid line). Gravity suppresses KHI above this length scale. Blue, green, magenta, and red circles mark the sizes and radii of S1, S2, S3, and a ghost bubble candidate, respectively.}
\end{figure}

Sloshing cold fronts are found to be subsonic in simulations and the motion slows down towards smaller radii (ZuHone et al.\ 2010, 2011; Ascasibar \& Markevitch 2006). Roediger et al.\ (2011) found that the sloshing cold fronts propagate at $\mathcal{M}\sim0.1$ in a simulation tailored to the Virgo Cluster. 
We note that there is no offset between the X-ray and the infrared centroids (Figure~\ref{fig:img2}-bottom-right), disfavoring a recent sloshing event in Fornax. 
We assume that this innermost front is 30--60\,Myr old, that is the sloshing time scale at $r=3$\,kpc (Figure~\ref{fig:sloshing}). This corresponds to an average sloshing velocity of 50--100\,km\,s$^{-1}$ ($\mathcal{M}\lesssim0.2$). 
Here, we conservatively adopt $\mathcal{M}<0.5$ (ZuHone et al.\ 2011).
The upper limits on ${f_{\nu}}$ and $B$ as a function of $\mathcal{M}$ are presented in Figure~\ref{fig:khi}. A full Spitzer viscosity can be ruled out and the strength of an ordered magnetic field should be smaller than 10$\mu$G.
However, as demonstrated in ZuHone et al.\ (2013), the suppression due to magnetic draping is only partial since 
the field lines may not be perfectly aligned.
{

More evident $\lesssim 1$\,kpc structures are present to the east at a similar radius (green and magenta boxes in Figure~\ref{fig:img2}). In particular, S3 has a best-fit temperature similar to that of the gas within the interface and is in pressure equilibrium with the external gas. 
We note that the temperature gradient to the east is much smoother than that in the west (Figure~\ref{fig:inn}). {This difference may be due to more evident KHI to the east, which may have disrupted a previously existing cold front.} KHI accelerates the turbulent mixing of gas of different phases, as we expect from theory and simulations (ZuHone et al.\ 2011; Roediger et al.\ 2015a,b). 
Large KHI rolls (above certain wavelengths, $\ell_{\rm max}$) are expected to be suppressed by gravity (Chandrasekhar 1961). We do not expect to observe KHI rolls larger than 

\footnotesize
\begin{equation}
\ell_{\rm max}\approx{\rm 18\,kpc}\left(\frac{\rho_c/\rho_h}{1.5}\right)^{-1}\left(\frac{U}{\rm 200\,km\,s^{-1}}\right)^2
\left(\frac{g}{3\times10^{-8}\rm cm\,s^{-2}}\right)^{-1},
\end{equation}
\normalsize

where $U$ is the shear velocity at the front and $g$ the gravitational acceleration (Roediger et al.\ 2012a). We assume $\rho_c=1.5\rho_h$ and $U(r)=0.3c_s(r)$; 
$c_s$ and $g(r)$ were calculated from the best-fit deprojected temperature profile and the X-ray hydrostatic mass profile (Paper~1).
We present the maximum KHI length scale as a function of radius at the cluster center in Figure~\ref{fig:khi2}. The size of these features stays below this upper limit. A potential ghost bubble of $\gtrsim$ 5\,kpc diameter resides at $r=13$\,kpc ($r=10$\,kpc in projection) to the northwest (Figure~\ref{fig:img}, see Paper 1 for details). Interestingly, Walker et al.\ (2017) propose that such giant X-ray surface brightness decrement may be due to KHI instead of AGN bubbles. The position and size of this ghost bubble candidate are also in the allowed parameter space for a giant KHI roll. 

Alternatively, S1, S2 and S3 may be thermally-unstable X-ray filaments resulting from gas cooling (David et al.\ 2014, 2017). The growth time of KHI can be estimated from (Roediger et al.\ 2012b)

\footnotesize
\begin{equation}
\tau_{\rm KH}=\frac{\sqrt{\Delta}}{2\pi}\frac{\ell}{U}=3.9\,{\rm Myr}\sqrt{\Delta}\frac{\ell}{\rm 10 kpc}\left(\frac{U}{\rm 400\,km\,s^{-1}}\right).
\end{equation}
\normalsize
We calculate $\tau_{\rm KH}$ for the maximum KHI length scale, the maximum growth time, at each radius as shown in Figure~\ref{fig:sloshing}. The maximum growth time of KHI is shorter than the cooling time by two orders of magnitude. KHI is thus somewhat favored as their origin. 
The time scale of KHI is also significantly shorter than the dynamical time scale (free fall time or buoyant rising time), implying that KHI, coupled with gas sloshing, may have played an important role in regulating the thermal state of the hot plasma at cluster centers. 
}

\section {\bf Conclusions}

The Fornax Cluster is a nearby low-mass cool-core cluster.  It displays prominent sloshing structure wrapping around the cluster center. 
We analyzed a mosaic {\sl XMM-Newton} observation and a total of 250\,ksec {\sl Chandra} observations centered on its BCG NGC~1399. We present the properties of the hot gas over a radial range of 1--250\,kpc.  
We found that: 

 \begin{itemize}

   \item Four sloshing fronts are visible at radii of 200\,kpc, 30\,kpc, 10\,kpc, and 3\,kpc respectively, allowing us to probe the merger history of Fornax. We speculate that all of these fronts were initiated by the infall of NGC~1404.

    \item The sloshing cold front to the northeast at a radius of 10\,kpc is the most prominent front, separating the low-entropy high-metallicity gas and the high-entropy low-metallicity gas.

    \item The spatial distribution of enriched ICM traces the spiral morphology of the gas sloshing. In this particular cluster, gas sloshing is more effective at redistributing the enriched gas than is AGN outflow.  
 \smallskip

    \item The innermost sloshing cold front resides within a radius of 3\,kpc which is at a smaller radius than the locations of AGN bubbles. A sub-kpc bright feature consistent with KHI eddies is visible at the interface which allows us to rule out a full Spitzer viscosity and put an upper limit of 10\,$\mu$G on the ordered magnetic field. We identify two other such features to the east where the temperature gradient is smaller, implying that KHI accelerates the heat transport. Alternatively, these features are thermally-unstable X-ray filaments. 
  
 \item The sloshing timescale is $10\times$ shorter than the cooling time at cluster center. Gas sloshing can be a plausible mechanism to alleviate the cooling problem. 

\end{itemize}

\section{\bf Acknowledgments}
We acknowledge helpful discussions with Eugene Churazov.
This work was supported by Chandra Awards GO1-12160X and GO2-13125X issued by the {\sl Chandra} X-ray Observatory Center which is operated by the Smithsonian Astrophysical Observatory under NASA contract NAS8-03060.

\bigskip

\begin{appendices}
\section{XMM-Newton Background modeling}

We consider two sources of background: Astrophysical X-ray background (AXB) and Non-X-ray background (NXB).  
The model of AXB contains a power law {\tt pow}$_{\rm CXB}$ with index $\Gamma=1.41$ characterizing the cosmic X-ray background (CXB, De Luca \& Molendi 2004), a thermal emission {\tt apec}$_{\rm MW}$ with a temperature of 0.2\,keV for the Milky Way emission, and another thermal emission {\tt apec}$_{\rm LB}$ with a temperature of 0.08\,keV for the Local Bubble emission (Smith et al.\ 2001). Metal abundance and redshift were fixed at 1 and 0 respectively for {\tt apec}$_{\rm MW}$ and {\tt apec}$_{\rm LB}$. All the AXB components except {\tt apec}$_{\rm LB}$ are expected to be absorbed by foreground (Galactic) cooler gas, characterized by the {\tt phabs} model. The best-fit surface brightnesses of all AXB components are listed in Table~A3, determined with a joint-fit of six offset pointings. 

The NXB component contains a set of fluorescent instrumental lines and a continuum spectrum for each MOS and pn detector. 
Fluorescent instrumental lines produced by the hard particles were modeled by a set of gaussian lines. Their centroid energies are listed in Table~A2, taken from Mernier et al.\ (2015) which were modified from Snowden \& Kuntz (2013). We set an upper limit of 0.3\,keV on each line width. We use a powerlaw model to characterize the continuum particle background component (Snowden \& Kuntz 2013). Even after filtering soft flare events from raw datasets, a
quiescent level of soft proton flare may remain. To estimate its effect on each CCD, 
we compared the area-corrected count rates in the 6--12\,keV energy band within the field of view (excluding the central 10$^{\prime}$) and that in the unexposed corners. 
If this ratio is below 1.15, we consider the event not contaminated by the residual soft proton flare (De Luca \& Molendi 2004). For contaminated events, we add an additional power law {\tt pow}$_{\rm SB}$ to model this residual soft proton component. Its index is allowed to vary between 0.1 and 1.4 (Snowden \& Kuntz 2013). The spectral fit was restricted to the 0.3--10.0\,keV energy band.
\newpage

\setcounter{table}{0}
\renewcommand*\thetable{A.\arabic{table}}
\begin{deluxetable*}{cccccccc}

\tablewidth{0pc}
 \centering
\tablecaption{XMM-Newton Observational log of the Fornax Cluster}
\tablehead{
\colhead{Obs ID}&\colhead{Name}&\colhead{Obs-Date}&\colhead{Exposure (ksec)$^*$}&\colhead{RA (J2000)}&\colhead{Dec (J2000)}&\colhead{Offset($^{\prime}$)}&\colhead{PI}}
\startdata
\hline
0012830101&NGC~1399&2001-06-27&29 (4, 4, 3)&03 38 29.30&-35 27 01.0& 0.03&Buote\\
0055140101&LP 944-20&2001-01-07&51 (42, 43, 40)&03 39 34.60&-35 25 51.0&13.41&Martin\\
0140950201&NGC~1386&2002-12-29&17 (16, 16, 14)&03 36 45.40&-35 59 57.0&39.04&Guainazzi\\
0210480101&RX J0337-3457&2005-01-04&49 (43, 44, 41)&03 37 24.70&-34 57 29.0&32.33&Stanford\\
0304940101&NGC~1404&2005-07-30&55 (22, 13, 17)&03 38 51.92&-35 35 39.8&9.81&Matsushita\\
0400620101&NGC~1399&2006-08-23&130 (81, 90, 51)&03 38 29.10&-35 27 03.0&0.03&Paerels\\
0550930101&Fornax offset A&2008-06-28&14 (11, 11, 8)&03 39 02.40&-35 01 55.2&26.02&Matsushita\\
0550930201&Fornax offset B&2008-06-27&17 (8, 8, 4)&03 39 26.16	&-34 49 37.2&39.20&Matsushita\\
0550930301&Fornax offset C&2008-07-17&17 (11, 12, 9)&03 40 27.12&-34 59 16.8&36.77&Matsushita\\
0550930401&Fornax offset D&2009-02-09&19 (14, 15, 13)&03 41 24.96&-35 10 04.8&39.70&Matsushita\\
0550930501&Fornax offset E&2009-02-23&19 (18, 18, 17)&03 41 40.80&-35 22 51.6&39.30&Matsushita\\
0550930601&Fornax offset F&2009-02-24&21 (18, 18, 15)&03 41 35.04&-35 37 48.0&39.35&Matsushita\\
0550930701&Fornax offset G&2009-02-24&19 (2, 4, 1)&03 40 52.08	&-35 50 02.4&37.07&Matsushita\\
0550931001&Fornax offset J&2008-06-25&22 (18, 19, 14)&03 41 40.80&-35 22 51.6&39.30&Matsushita\\
0550931201&Fornax offset L&2008-06-25&13 (8, 8, 4)&03 36 15.60&-35 32 56.4&27.79&Matsushita\\
0550931401&Fornax offset N&2008-06-26&12 (11, 11, 9)&03 37 11.52&-35 17 34.8&18.41&Matsushita\\
0694670101&	NGC 1380 ULX&2013-01-25&103 (69, 70, 59)&03 36 25.20&	-34 59 18.0&37.53&Sarazin
\enddata
\tablecomments{*Effective exposure times of MOS1, MOS2, and pn are listed in the brackets.}
\end{deluxetable*}

 \begin{deluxetable}{cc|cc}

\tablewidth{0pc}
 \centering
\tablecaption{Fluorescent instrumental lines produced by the hard particles}
\tablehead{
\colhead{Energy (keV)}&\colhead{Line}&\colhead{Energy (keV)}&\colhead{Line}}
\startdata
\multicolumn{2}{c}{MOS} &\multicolumn{2}{c}{pn} \\
\hline
1.49&Al K$_{\alpha}$&1.48&Al K$_{\alpha}$\\
1.75&Si K$_{\alpha}$&4.51&Ti K$_{\alpha}$\\
5.41&Cr K$_{\alpha}$&5.42&Cr K$_{\alpha}$\\
5.90&Mn K$_{\alpha}$&6.35&Fe K$_{\alpha}$\\
6.40&Fe K$_{\alpha}$&7.47&Ni K$_{\alpha}$\\
7.48&Ni K$_{\alpha}$&8.04&Cu K$_{\alpha}$\\
8.64&Zn K$_{\alpha}$&8.60&Zn K$_{\alpha}$\\
9.71&Au L$_{\alpha}$&8.90&Cu K$_{\beta}$\\
&&9.57&Zn K$_{\beta}$
\enddata
\end{deluxetable}

 \begin{deluxetable}{ccc}

\tablewidth{0pc}
 \centering
\tablecaption{X-ray Background model parameters}
\tablehead{
\colhead{Background}&\colhead{kT (keV) or $\Gamma$}&\colhead{norm ($\times10^{-6}$ per arcmin$^{2}$)}}
\startdata
CXB&1.41&$0.61\pm0.07$\\
MW&0.2&$0.44\pm0.07$\\
LB&0.08&$1.33\pm0.11$
\enddata
\end{deluxetable}

\end{appendices}

\end{document}